\documentclass[prd,12pt,nofootinbib,preprint]{revtex4}
\usepackage[T1]{fontenc}
\usepackage{amsmath,amssymb}
\usepackage{epsfig}
\usepackage{url}
\usepackage{subfigure}
\usepackage{graphicx}
\usepackage{grffile}
\usepackage[usenames,dvipsnames]{color}
\usepackage{slashed}
\usepackage{array}
\usepackage{multirow}
\usepackage[colorlinks,citecolor=blue]{hyperref}
\usepackage{color}
\usepackage{cancel}
\usepackage{gensymb}

\def\a {\alpha}

\def\l {\lambda}

\def\bar {\overline}

\def\be {\begin{equation}}
\def\ee {\end{equation}}
\def\beq {\begin{equation}}
\def\eeq {\end{equation}}
\def\bea {\begin{eqnarray}}
\def\eea {\end{eqnarray}}

\newcommand{\besub}{\begin{subequations}}
\newcommand{\eesub}{\end{subequations}}

\def\beq{\begin{equation}}
\def\eeq{\end{equation}}
\def\barr{\begin{array}}
\def\earr{\end{array}}

\begin{document}
\title{Type-I thermal leptogenesis in $Z_3$-symmetric three Higgs doublet model}

\author{Indrani Chakraborty}
\email{indranic@iitk.ac.in, indrani300888@gmail.com}
\affiliation{Department of Physics, Indian Institute of Technology Kanpur, Kanpur, Uttar Pradesh-208016, India}
 
\author{Himadri Roy}
\email{himadrir@iitk.ac.in, himadri027roy@gmail.com}
\affiliation{Department of Physics, Indian Institute of Technology Kanpur, Kanpur, Uttar Pradesh-208016, India} 
\begin{abstract} 
Our present work explores the possibility of neutrino mass generation through {\em Type-I see-saw} mechanism and provides an explanation of the baryon asymmetry of the Universe via thermal leptogenesis in the framework of a $Z_3$-symmetric three Higgs doublet model (3HDM) augmented with three right-handed neutrinos. Here the thermal leptogenesis is initiated by the out-of-equilibrium decay of the lightest heavy neutrino $N_1$. The constraints arising out of the scalar sector put strong bound on the model parameter $\tan \beta$, which in turn takes part in the computation of the lepton asymmetry $\epsilon$. Lepton asymmetry being converted partially into the baryon asymmetry by electroweak sphelaron processes, will account for the required baryon asymmetry satisfying the current data. We therefore analyse the parameter space consistent with the constraints arising from neutrino oscillation, lepton asymmetry and baryon asymmetry together, last one turning out to be the most stringent one.
\end{abstract} 
\maketitle
\section{Introduction}
\label{intro}
While the Standard Model (SM) particle spectrum is complete after Higgs discovery \cite{Aad:2012tfa,Chatrchyan:2012xdj}, there are several reasons to believe that this is not the ultimate story but only an effective theory valid up to some high energy scale, above which some other theory takes over. Like other shortcomings of SM, explanation of  neutrino mass generation necessitates a beyond-the-SM (BSM) scenario incorporating different types of see-saw mechanisms \cite{Minkowski:1977sc,Yanagida:1979as,Mohapatra:1979ia,GellMann:1980vs,Glashow:1979nm,Schechter:1980gr}. Addition of right-handed (RH) neutrinos, which are singlets under SM gauge group, results in the  creation of neutrino mass via {\em Type-I see-saw} mechanism \cite{Minkowski:1977sc,Yanagida:1979as,Mohapatra:1979ia,GellMann:1980vs}. Besides, the observed imbalance between the number of baryons and antibaryons is yet another issue that remains unaddressed within the ambit of SM \cite{Rubakov:1996vz,Morrissey:2012db}. The dynamic generation of {\em baryon asymmetry} needs to comply with three Sakharov conditions \cite{Sakharov:1967dj}, which require : (a) baryon number violation, (b) $C$ or $CP$-violation, (c) out-of-equilibrium condition. The current data reads \cite{Aghanim:2018eyx} :
\be
\eta_B \equiv \frac{n_B - n_{\bar{B}}}{n_\gamma} = (6.12 \pm 0.04) \times 10^{-10} \,.
\label{Bar_asy}
\ee
$n_B, ~ n_{\bar{B}}, ~ n_\gamma$ being number densities of baryons, anti-baryons and photons respectively. Thus new physics (NP) needs to be introduced to compensate the due amount of baryon asymmetry within SM.

The out-of-equilibrium decay of the RH heavy neutrinos in {\em Type-I see-saw} mechanism, induces {\em leptogenesis} \cite{Fukugita:1986hr,Davidson:2008bu}, that can make up for the aforementioned baryon imbalance. Complex Yukawa couplings give rise to $CP$-violation, thereby fulfilling the required criteria for generating baryon asymmetry. At the epoch of generation of asymmetry, the decay rate being slower than the expansion rate of the Universe, out-of-equilibrium condition is automatically fulfilled. Finally a partial conversion of lepton asymmetry (created during the out-of-equilibrium decay of heavy neutrinos) to baryon asymmetry, occurs through $(B+L)$ violating electroweak (EW) {\em sphelaron processes} \cite{Kuzmin:1985mm}.  

As stated earlier, the inability of SM, to address the issues of  neutrino mass generation and baryon asymmetry calls for a BSM scenario. A particular way in this direction is to extend the SM by spin-0 degrees of freedom only. Moreover, extending the SM scalar sector by $SU(2)_L$ doublets only, is an attractive choice since the tree level electroweak $\rho$-parameter is kept intact. The most minimal multi-doublet extension comprises two Higgs doublets leading up to what is known as two Higgs doublet models (2HDM) \cite{Branco:2011iw, Bhattacharyya:2015nca}.  However, as there is no fundamental principle to pinpoint the exact number of doublets present, more non-minimal extensions are also possible. In fact, three Higgs doublet models 
(3HDMs) \cite{Ferreira:2008zy,Machado:2010uc,Aranda:2012bv,Ivanov:2012ry,Ivanov:2012fp,Felipe:2013ie,Felipe:2013vwa,Keus:2013hya,Das:2014fea,Ivanov:2014doa,Maniatis:2014oza,Chakrabarty:2015kmt,Merchand:2016ldu,Emmanuel-Costa:2016vej,Bento:2017eti,Emmanuel-Costa:2017bti,Pramanick:2017wry,Varzielas:2015joa,Moretti:2015tva,Maniatis:2015kma,Moretti:2015cwa,Keus:2014jha,deMedeirosVarzielas:2019rrp,Camargo-Molina:2017klw,Das:2019yad} have been attracting attention for quite some time now. The main motivation of 3HDM lies in the fact that, the masses and mixings of three fermionic generations can be properly reproduced, when these three doublets are connected to the three fermionic generations via appropriate symmetries. Examples of such discrete symmetries include $A_4, S_4, S_3, Z_3$ etc. A $Z_3$-symmetric 3HDM \cite{Bento:2017eti,Ferreira:2008zy,Das:2019yad} resembles the {\em  democratic} 3HDM (where three doublets individually couple to up-type quarks, down-type quarks and leptons) \cite{Cree:2011uy,Akeroyd:2018axd,Akeroyd:2019mvt} via proper $Z_3$-charge assignment to quarks and leptons as will be discussed later. Another important aspect of this $Z_3$-symmetric 3HDM is to promote the {\em natural flavour conservation (NFC)} by prohibiting the tree level flavour changing neutral currents (FCNCs).

In this paper, we uphold the $Z_3$-symmetric 3HDM augmented with three heavy RH neutrinos as a possible framework to address the two aforementioned shortcomings of SM. In particular, here we shall focus on  {\em thermal leptogenesis} \cite{Chianese:2018rnq,Hugle:2018qbw,Ipek:2018sai,Ibe:2016gir,Ishihara:2015uua,DiBari:2015oca,Davidson:2008bu, Biswas:2018sib}, which allows hierarchical heavy neutrino masses, mass of one of them being much smaller than others. Besides, only thermal generation and out-of-equilibrium decay of lightest heavy neutrino will play the crucial role in generating lepton asymmetry. In presence of three RH neutrinos, mass generation of light neutrinos will be possible  via {\em Type-I see-saw} mechanism. As can be seen later, the entire parameter space will be constrained by the restrictions coming from the scalar sector, as well as the more stringent constraints arising from neutrino oscillation data, lepton asymmetry and baryon asymmetry respectively. Thermal type-I leptogenesis in a minimal scenario containing RH neutrinos along with SM Higgs doublet has been analysed in \cite{Buchmuller:2002rq,Giudice:2003jh,Buchmuller:2004nz}. Natural and thermal leptogenesis has been studied in the framework of 2HDM extended by RH neutrinos in \cite{Clarke:2015hta,Atwood:2005bf}. Analysis of the scalar sector of the $Z_3$-symmetric 3HDM along with three heavy RH neutrinos has not been performed earlier in light of the theoretical, experimental constraints. In addition,  any study of type-I thermal leptogenesis has not been done within this particular model. Thus there is a huge impulse for analysing this model in light of type-I thermal leptogenesis. There can be another variant of leptogenesis, in which the $CP$-asymmetry is enhanced by considering the mass-splitting between any two of the heavy neutrinos to be comparable with their decay width. This type of leptogenesis is termed as {\em Resonant leptogenesis} \cite{Pilaftsis:2003gt,Dev:2017wwc}. Since the lower bound on the heavy neutrino mass is relaxed in this case, collider searches involving these neutrinos are feasible in the future colliders. To understand the importance of flavor effects on leptogenesis, we refer the readers to go through the papers \cite{Abada:2006fw, Nardi:2006fx, Abada:2006ea, Blanchet:2006be, Dev:2017trv, Samanta:2019yeg}. However we shall restrict ourselves in studying {\em thermal leptogenesis} in this paper and shall not consider the other variants.

This paper is structured as follows. Sec. \ref{sec:1} contains the information regarding the particle content of the model considered for analysis. Sec. \ref{sec:2} comprises of detailed discussion of several constraints imposed on the parameter space. In sec. \ref{sec:3}, we elaborate the fitting of neutrino oscillation data using Casas Ibarra parametrization. Sec.\ref{sec:4} deals with thermal leptogenesis, {\em i.e.} solutions of Boltzman equations. In sec. \ref{sec:5}, we present analysis and results. Finally we summarize and conclude in sec. \ref{sec:6}.
\section{Model}
\label{sec:1}
In this analysis, we consider $Z_3$-symmetric 3HDM comprising of three $SU(2)_L$ doublets $\phi_1, \phi_2$ and $\phi_3$ each with hyper-charge $Y= +1$ \footnote{We have calculated the hyper-charge $Y$ by using the relation : $Q = T_3 + \frac{Y}{2}$, $T_3$ and Q being the weak isospin and electric charge.}, augmented with three heavy RH neutrinos $N_{1R}$, $N_{2R}$, $N_{3R}$. For simplicity, we shall denote these three heavy neutrinos as $N_1, N_2, N_3$ throughout the analysis. The complete description of different sets of quantum numbers assigned to all the particles can be found in table \ref{tab:1}.

\begin{table}[htpb!]
\begin{center}
\begin{tabular}{|c|c|c|c|c|}
\hline
Fields & $SU(2)_L$ &$SU(3)_C$ &  \hspace{2mm}$Z_3$ \hspace{2mm} &\hspace{2mm} $Y$ \hspace{2mm}\\
\hline
\hline
$\phi_1$ & 2& 1& $\omega$ & +1\\
\hline
$\phi_2$ &2 &1 & $\omega^2$ & +1 \\
\hline
$\phi_3$ & 2 & 1 & 1 & +1 \\
\hline
$Q_L$ &2 &3 & $\omega$ & $+\frac{1}{3}$\\
\hline
 $u_R$ &1 & 3& $\omega^2$ & $+\frac{4}{3}$\\
 \hline 
 $d_R$ &1 & 3& $\omega^2$ & $-\frac{2}{3}$\\
 \hline
$L_L$ & 2& 1& 1 & $-1$\\
\hline
$l_R$ & 1& 1& 1 & $-2$\\ 
\hline
$N_{iR}, i =1,2,3$ &1 & 1& 1& 0\\
\hline
\end{tabular}
\end{center}
\caption{Different quantum numbers assigned to the particles in the model. Here $\omega = e^{\frac{2 i \pi}{3}}$.}
\label{tab:1}
\end{table}

\subsection{$Z_3$-symmetric Scalar Lagrangian}
\label{subsec:1a}
Following the quantum numbers assigned to the doublets, as mentioned in table \ref{tab:1}, the $Z_3$-symmetric scalar potential involving $\phi_1, \phi_2$ and $\phi_3$ can be written as \cite{Das:2019yad},
\bea
V(\phi_1, \phi_2, \phi_3) &=& m_{11}^2 (\phi_1^\dag \phi_1) + m_{22}^2 (\phi_2^\dag \phi_2) + m_{33}^2 (\phi_3^\dag \phi_3) \nonumber \\
&& + \frac{\lambda_1}{2} (\phi_1^\dag \phi_1)^2 + \frac{\lambda_2}{2} (\phi_2^\dag \phi_2)^2 + \frac{\lambda_3}{2} (\phi_3^\dag \phi_3)^2 \nonumber \\
&& + \lambda_4 (\phi_1^\dag \phi_1)(\phi_2^\dag \phi_2) + \lambda_5 (\phi_1^\dag \phi_1)(\phi_3^\dag \phi_3) + \lambda_6 (\phi_2^\dag \phi_2)(\phi_3^\dag \phi_3) \nonumber \\
&& + \lambda_7 (\phi_1^\dag \phi_2)(\phi_2^\dag \phi_1) + \lambda_8 (\phi_1^\dag \phi_3)(\phi_3^\dag \phi_1) + \lambda_9 (\phi_2^\dag \phi_3)(\phi_3^\dag \phi_2) \nonumber \\
&& + \left[\lambda_{10} (\phi_1^\dag \phi_2)(\phi_1^\dag \phi_3) + \lambda_{11} (\phi_1^\dag \phi_2)(\phi_3^\dag \phi_2) + \lambda_{12} (\phi_1^\dag \phi_3)(\phi_2^\dag \phi_3)+ {\rm h.c.}\right]
\label{scalar-pot}
\eea
After symmetry breaking, $\phi_i$ can be expressed as,
\begin{eqnarray}
\phi_i = \begin{pmatrix}
h_i^+ \\
\frac{1}{\sqrt{2}} (v_i + h_i + i \rho_i)
\end{pmatrix}, i=1,2,3
\end{eqnarray}
$v_i$ being the vacuum expectation value (VEV) of $\phi_i$. Two important parameters of the model $\tan \beta $ and $\tan \gamma$ can be expressed as the ratios of VEVs of doublets :
$\tan\beta = \frac{\sqrt{v_1^2 + v_2^2}}{v_3},
\tan\gamma = \frac{v_2}{v_1}$. Therefore $v_1, v_2$ and $v_3$ can be written in terms of the mixing angles $\beta$ and $\gamma$ as :
\bea
v_1 &=& v \sin \beta~\cos \gamma , \nonumber \\
v_2 &=& v \sin \beta ~ \sin \gamma , \nonumber \\
 v_3 &=& v \cos \beta \,, ~{\rm with}~ v = \sqrt{v_1^2 + v_2^2 + v_3^2 } = 246~ {\rm GeV}
\eea

The quartic couplings $[\lambda_1, \lambda_2,... \lambda_{12}]$ and the doublet VEVs $[v_1, v_2, v_3]$ are taken to be real to avoid any kind of $CP$-violation in the scalar potential. The particle spectrum of the model comprises of seven physical scalars, namely $h, H_1, H_2, A_1, A_2, H_1^{\pm}, H_2^{\pm}$. Twelve quartic couplings $[\lambda_1, \lambda_2,... \lambda_{12}]$ can be rewritten in terms of the aforementioned seven physical masses ( $M_h, M_{H_1}, M_{H_2}, M_{A_1}, M_{A_2}, M_{H_1^{\pm}}, M_{H_2^{\pm}}$) and five mixing angles, {\em i.e.} three in the $CP$-even sector ($\alpha_1, \alpha_2, \alpha_3$), one in $CP$-odd sector ($\gamma_1$) and one in charged scalar sector ($\gamma_2$) \cite{Das:2019yad}. The lightest neutral physical state $h$ resembles the SM Higgs boson with mass 125 GeV at the {\em Alignment limit} defined as : $\alpha_1 = \gamma , \alpha_2 + \beta = \frac{\pi}{2}$ \cite{Das:2019yad}.

The details of the scalar sector of $Z_3$-symmetric 3HDM including the basis transformations from flavor basis to mass basis etc. can be found in \cite{Das:2019yad}. To avoid repetition, we shall not provide the same details here. 
\subsection{Yukawa Lagrangian}
\label{subsec:1b}
Due to the particular assignment of $Z_3$-charges to the fields (shown in table \ref{tab:1}), the flavor changing neutral currents (FCNCs) are forbidden in this model. Up-type, down-type quarks and leptons will acquire masses through the couplings with $\phi_1, \phi_2$ and $\phi_3$ respectively. Due to the presence of three heavy RH neutrinos, SM light neutrinos can also acquire masses via Type-I see-saw mechanism, only $\phi_3$ being responsible for the mass generation of neutrinos.

Thus we can write down the $Z_3$-symmetric Yukawa Lagrangian along with the Majorana mass terms for the heavy neutrinos as :
\bea
-\mathcal{L}_Y &=& y_1 \bar{L}_1 \tilde{\phi_3} N_1 + y_2 \bar{L}_1 \tilde{\phi_3} N_2 + y_3 \bar{L}_1 \tilde{\phi_3} N_3 
  \nonumber \\ 
&& + y_4 \bar{L}_2 \tilde{\phi_3} N_1 + y_5 \bar{L}_2 \tilde{\phi_3} N_2 + y_6 \bar{L}_2 \tilde{\phi_3} N_3  \nonumber \\
&& + y_7 \bar{L}_3 \tilde{\phi_3} N_1 + y_8 \bar{L}_3 \tilde{\phi_3} N_2 + y_9 \bar{L}_3 \tilde{\phi_3} N_3 \nonumber \\
&& + \frac{1}{2} M_1 \bar{N_1}^c N_1 +  \frac{1}{2} M_2 \bar{N_2}^c N_2 +  \frac{1}{2} M_3 \bar{N_3}^c N_3 + {\rm h.c.}
\label{Yuk:lag}
\eea
Here $L_i$ are left-handed (LH) lepton doublets and $\tilde{\phi_i} = i \sigma_2 \phi_i^{*}$. As mentioned earlier, only $\phi_3$ will be responsible for generating SM light neutrino masses. Yukawa couplings $y_j$ are taken to be complex for generating $CP$-asymmetry in leptogenesis. The real and imaginary parts of the Yukawa couplings $y_j$ are constrained by recent neutrino oscillation data \cite{Esteban2019}, as will be discussed elaborately in section \ref{sec:3}.

\section{Constraints to be considered}
\label{sec:2} 
\begin{figure}[htpb!]
\includegraphics[scale=0.65]{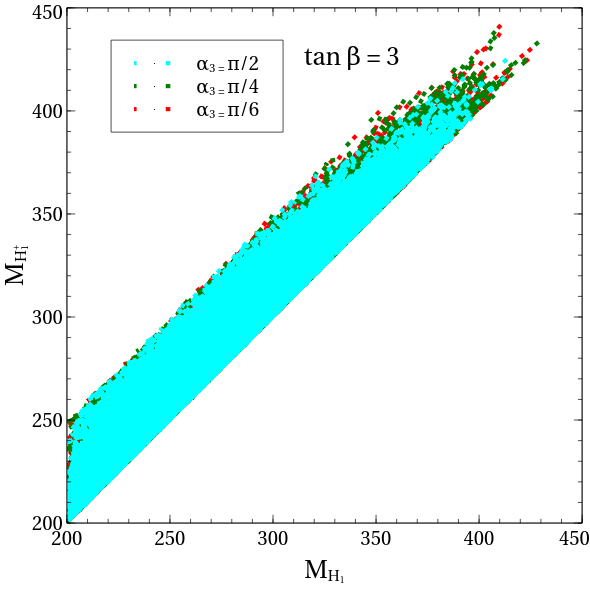}
\caption{Plot of the parameter space in $M_{H_1}$ vs. $M_{H_1^+}$ plane for $\alpha_3 = \frac{\pi}{6}, \frac{\pi}{4}, \frac{\pi}{2}$ and $\tan \beta = 3$.}
\label{fig:1}
\end{figure}

For the analysis, we shall consider a multi-dimensional parameter space, spanned by the following independent parameters : $\tan\beta , \gamma, \gamma_1 , \gamma_2, M_h,  M_{H_1}, M_{H_2}, M_{A_1}, M_{A_2}, M_{H_1^{\pm}}, M_{H_2^{\pm}}, \alpha_3$ ($\alpha_1, \alpha_2$ are connected to $\gamma_1$ and $\gamma_2$ respectively through the {\em alignment} conditions : $\alpha_1 = \gamma, ~ \alpha_2 + \beta = \frac{\pi}{2}$, thus are not independent).  Since {\em Alignment limit} will be imposed strictly, the lightest Higgs $h$ being SM-like, $M_h$ is taken to be 125 GeV. We have checked that the variation of $\alpha_3$ hardly induces any change in the parameter space. The effect of scanning over the other variables like physical masses and mixing angles surpasses the mild effect of changing $\alpha_3$. To illustrate this, we refer to the plot of $M_{H_1}$ vs. $M_{H_1^+}$ plane for $\tan \beta = 3$ and three different values of $\alpha_3 = \frac{\pi}{6}, \frac{\pi}{4}, \frac{\pi}{2}$ in fig.\ref{fig:1}. It shows that for three different values of $\alpha_3$, the parameter space in the mass plane changes only mildly. The same conclusion can be drawn for the other masses and $\tan \beta$ as well. Therefore, to
simplify the numerical scans, we fix $\alpha_3 = \frac{\pi}{4}$ throughout the rest of the analysis. 

In addition to these, the constraints to be imposed on the parameter space are discussed below.
\subsection{Theoretical constraints}
\label{subsec:2a}
\begin{itemize}
\item The real quartic couplings $[\lambda_1, \lambda_2,... \lambda_{12}]$ are taken to be pertubative , {\em i.e.} $|\lambda_i| \leq 4\pi$. 
\item Yukawa couplings $y_j$ ($j = 1,2,...9$) are constrained from the neutrino oscillation data and constraints arising from leptogenesis, as will be discussed later. There is also an upper bound of $|y_j| \leq \sqrt{4\pi}$ arising from perturbativity.
\item Boundedness of the scalar potential (eq.(\ref{scalar-pot})) can be ensured by satisfying following stability conditions involving the quartic couplings :
\begin{enumerate}
\item $\lambda_1, \lambda_2, \lambda_3 \geq 0$
\item $\lambda_4 \geq - \sqrt{ \lambda_1 \lambda_2}$
\item $\lambda_5 \geq - \sqrt{ \lambda_1 \lambda_3}$ 
\item $\lambda_6 \geq - \sqrt{ \lambda_2 \lambda_3}$
\item $\lambda_5 + \lambda_8  \geq - \sqrt{\lambda_1 \lambda_3}$
\item $\lambda_6 + \lambda_9  \geq - \sqrt{\lambda_2 \lambda_3}$
\item $\lambda_4 + \lambda_7  \geq - \sqrt{\lambda_1 \lambda_2}$
\item $|\lambda_{10}|, |\lambda_{11}|, |\lambda_{12}| < |\lambda_i|,~ i=1,2,...,9$.
\end{enumerate}
Here first seven conditions come from the phase-invariant part of the scalar potential which includes all the terms in eq.(\ref{scalar-pot}) except last three terms. The last stability condition arises only from the $Z_3$-symmetry ensuring terms (last three terms in eq.(\ref{scalar-pot})).
\end{itemize}
\subsection{Constraints from oblique parameters}
\label{subsec:2b}
In presence of additional scalars in the model, the oblique parameters $S,T,U$ will be modified accordingly. The present
limits on their deviation from SM values are \cite{Tanabashi:2018oca}:

\bea
\Delta S &=& 0.02 \pm 0.01, \nonumber \\ 
\Delta T &=& 0.07 \pm 0.12, \nonumber \\
\Delta U &=& 0.00 \pm 0.09. 
\eea

Specially the $Z_3$-symmetric 3HDM parameter space is sensitive to the deviation of $T$-parameter from SM value, because this deviation controls the mass-splitting between the charged and the neutral scalars. We have ensured the compatibility with $T$-parameter constraint by keeping the mass-splitting between the charged and the neutral scalars $\sim$ 50 GeV. 
\subsection{Constraints on Higgs signal-strengths from LHC data}
\label{subsec:2c}
To make the parameter space compatible with the current LHC data, one has to compute Higgs signal strengths in different Higgs decay channels. For the decay channel $h \rightarrow X Y$, the signal strength $\mu_{X Y}$ can be computed as the ratio of cross section of Higgs production via $p-p$ collision times the branching ratio of Higgs decay into the channel $h \rightarrow X Y$ in 3HDM and the same quantity measured in the SM:

\bea
\mu_{X Y} = \frac{\sigma^{\rm{3HDM}}(pp \rightarrow h)~ {\rm BR^{3HDM}}(h \rightarrow X Y)}{\sigma^{\rm{SM}}(pp \rightarrow h)~ {\rm BR^{SM}}(h \rightarrow X Y)}.
\label{sig-str-1}
\eea

Among all Higgs production process, the dominant contribution at the LHC comes from the gluon-gluon fusion process mediated by heavy quarks in triangular loops. The parton-level cross section can be written as \cite{Djouadi:2005gi}

\bea
\sigma(gg \rightarrow h) = \frac{\pi^2}{8 M_h} \Gamma (h \rightarrow gg)~ \delta(\hat{s} - M_h^2),
\label{xsec:gg-h}
\eea
$\hat{s}$ being gluon-gluon invariant energy squared.

Using eqs. (\ref{sig-str-1}) and (\ref{xsec:gg-h}), one can rewrite the signal strength $\mu_{X Y}$ as :

\bea 
\mu_{X Y} &=& \frac{\sigma^{\rm{3HDM}}(gg \rightarrow h)}{\sigma^{\rm{SM}}(gg \rightarrow h)} ~\frac{\Gamma_{X Y}^{\rm{3HDM}}(h \rightarrow  X Y )}{\Gamma_{\rm{tot}}^{\rm{3HDM}}} ~\frac{\Gamma_{\rm{tot}}^{\rm{SM}}}{\Gamma_{X Y}^{\rm{SM}}(h \rightarrow X Y)}, \nonumber \\
&=& \frac{\Gamma^{\rm{3HDM}}(h \rightarrow gg)}{\Gamma^{\rm{SM}}(h \rightarrow gg)} ~\frac{\Gamma_{X Y}^{\rm{3HDM}}(h \rightarrow X Y)}{\Gamma_{\rm{tot}}^{\rm{3HDM}}} ~\frac{\Gamma_{\rm{tot}}^{\rm{SM}}}{\Gamma_{X Y}^{\rm{SM}}(h \rightarrow X Y)}.
\eea
where $\Gamma_{\rm{tot}}$ stands for the total decay width.

Since the {\em Alignment limit} is being invoked strictly, the lightest Higgs $h$ being SM like, the Higgs signal strengths in the $WW, ZZ, b \bar{b}, \tau^+ \tau^-$ mode are satisfied automatically. $\Gamma(h \rightarrow \gamma \gamma)$ receives an extra contribution coming from the charged Higgs mediated loop and are modified. At the exact {\em Alignment limit}, the total Higgs decay width coincides with that of the SM Higgs $h$. Thus the signal strength $\mu_{h \rightarrow \gamma \gamma}$ can be approximated to $\frac{\Gamma^{\rm{3HDM}}(h \rightarrow \gamma \gamma)}{\Gamma^{\rm{SM}}(h \rightarrow \gamma \gamma)}$. Expressions for the decay width  $\Gamma(h \rightarrow \gamma \gamma)$ can be found in appendix \ref{app : A}. We have used $2\sigma$-deviation from the allowed values of signal strength to scan the parameter space \cite{Aaboud:2018xdt}.
\section{Fitting of neutrino-data}
\label{sec:3} 
As mentioned earlier, the Yukawa couplings need to be complex in order to generate lepton asymmetry required for leptogenesis. Following the Yukawa Lagrangian in eq.(\ref{Yuk:lag}), after symmetry breaking, the Dirac mass matrix $M_D$ can be computed as :
\bea
M_D = \frac{v_3}{\sqrt{2}} ~ Y_{ij} = \frac{v_3}{\sqrt{2}}
\begin{pmatrix}
y_1 & y_2 & y_3 \\
y_4 & y_5 & y_6 \\
y_7 & y_8 & y_9
\end{pmatrix}
= \frac{v \cos \beta}{\sqrt{2}}
\begin{pmatrix}
(y_{1R} + i y_{1I}) & (y_{2R} + i y_{2I}) & (y_{3R} + i y_{3I}) \\
(y_{4R} + i y_{4I}) & (y_{5R} + i y_{5I}) & (y_{6R} + i y_{6I}) \\
(y_{7R} + i y_{7I}) & (y_{8R} + i y_{8I}) & (y_{9R} + i y_{9I})
\end{pmatrix}.
\label{MDirac}
\eea
Here complex Yukawa couplings $y_j$s are decomposed into real and imaginary parts as : $y_{jR}$
and $y_{jI}$ respectively. Majorana mass matrix $M_R$ is assumed to be diagonal for simplicity :
\bea
M_R = \begin{pmatrix}
M_1 & 0 & 0 \\
0 & M_2 & 0 \\
0 & 0 & M_3
\end{pmatrix}.
\eea
For the mass generation of light neutrinos through Type-I see-saw mechanism, the  neutrino mass matrix can be expressed in terms of Dirac mass matrix $M_D$ and Majorana mass matrix $M_R$ as : 
\bea
M_\nu = - M_D M_R^{-1} M_D^T \,.
\eea
$M_\nu$ can be diagonalised to get the light neutrino masses by the transformation : 
\bea 
U_{\rm PMNS}^T~ M_\nu ~U_{\rm PMNS} = {\rm diag}(m_1,m_2,m_3) = \widehat{M_\nu} \,.
\eea
 where $m_1, m_2, m_3$ are three light neutrino masses, $U_{\rm PMNS}$ is the Pontecorvo-Maki-Nakagawa-Sakata matrix (PMNS) matrix  and can be written as :
\bea
U_{\rm PMNS} = \begin{pmatrix}
c_{13}c_{12} & c_{13} s_{12} & s_{13}e^{-i \delta_{\rm CP}} \\
-s_{12}c_{23}-c_{12}s_{23}s_{13}e^{i \delta_{CP}} & c_{12} c_{23}-s_{12} s_{23} s_{13} e^{i \delta_{\rm CP}}& s_{23} c_{13} \\
s_{12} s_{23} - c_{12} c_{23} s_{13}e^{i \delta_{\rm CP}} & -c_{12} s_{23} - s_{12} c_{23} s_{13} e^{i \delta_{\rm CP}} & c_{23} c_{13}
\end{pmatrix},
\eea
where $c_{ij} \equiv \cos \theta_{ij} , s_{ij} \equiv \sin \theta_{ij}$ and $\delta_{CP}$ is the $CP$-violating phase.
  To parametrize the elements of $M_D$, one can use the parametrization proposed by Casas and Ibarra (CI) \cite{Xing:2009vb}, as will be mentioned in detail in the next subsection.
\subsection{Casas Ibarra Parametrization}
\label{subsec:3a}
According to the CI parametrization \footnote{Usually, the charged lepton mass matrix and $M_R$ are diagonal, real and positive in the basis in which CI parametrization is defined.}\cite{Xing:2009vb}, $M_D$ can be rewritten as : 
\bea
M_D = \frac{v_3}{\sqrt{2}} Y_{ij} = i~U_{\rm PMNS} \sqrt{\widehat{M_\nu}} ~\mathcal{O} ~\sqrt{M_R} \,.
\label{CI_param}
\eea
$\mathcal{O}$ being a general complex orthogonal matrix, with complex angles $\theta, \chi, \psi$, can be expressed as \cite{Davidson:2008bu},
 \begin{align}
\mathcal{O} =
\begin{pmatrix} 
c_{\chi} c_{\psi} & c_{\chi}s_{\psi} & s_{\chi} \\ 
-c_{\theta} s_{\psi} - s_{\theta} s_{\chi} c_{\psi} & c_{\theta} c_{\psi} - s_{\theta} s_{\chi} s_{\psi} & s_{\theta} c_{\chi} \\ 
s_{\theta} s_{\psi} - c_{\theta} s_{\chi} c_{\psi} & -s_{\theta} c_{\psi} - c_{\theta} s_{\chi} s_{\psi} & c_{\theta} c_{\chi}  
\end{pmatrix},
\label{ortho_mat}
\end{align}
where $c_\alpha , s_\alpha$ are the shorthand notations for $\cos \alpha$ and $\sin \alpha$ respectively. Here $\widehat{M_\nu}$ is the diagonal light neutrino mass matrix.

The angles $\theta, \chi, \psi$ in eq.(\ref{ortho_mat}), can be complex in general, but for our analysis, we have chosen the phase associated with the angles to be zero, {\em i.e.} the angles are chosen to be real for simplicity. One of the three light neutrinos is taken to be massless, {\em i.e.} $m_1 = 0$. We have considered the Normal hierarchy (NH) among $m_1, m_2, m_3$. From eq.(\ref{CI_param}), it is evident that one can evaluate the elements of matrix $M_D$, {\em i.e.} the real and imaginary parts of Yukawa couplings (18 real parameters), in terms of the elements of $U_{\rm PMNS}$ matrix (which is known from neutrino oscillation data), angles $\theta, \chi, \psi$ in orthogonal matrix $\mathcal{O}$ and $M_1, M_2, M_3$ in $M_R$. Complex $U_{\rm PMNS}$ matrix at the right hand side of eq.(\ref{CI_param}) in turn necessitates complex Yukawa couplings $y_j$s in $M_D$ at the left hand side of the same equation. Here we have solved the real and imaginary parts of the Yukawa couplings using Casas
Ibarra parametrization, in terms of the matrix elements of $U_{\rm PMNS}$, $M_\nu , \mathcal{O}, M_R$ and
$v_3$ following eq.(\ref{MDirac}) and eq.(\ref{CI_param}), to make them consistent with the neutrino oscillation data. Thus the real and imaginary parts of the Yukawa couplings pick up a $\beta$-
dependence. Discussions regarding this will be elaborated  in section \ref{sec:4}.

\section{Leptogenesis}
\label{sec:4}
During this analysis, we aim to explore that portion of the parameter space, where the model parameters satisfy the constraints coming from neutrino oscillation data, as well as the current bound on baryon asymmetry. The main mechanism of generating baryon asymmetry here is {\em leptogenesis}, through which the lepton asymmetry is produced. In this scenario, the lepton asymmetry is originated by the $CP$-violating, out-of-equilibrium decay of the lightest heavy RH Majorana neutrino $N_1$. 
In the limit of hierarchical neutrino masses, {\em i.e.} $M_1 << M_2, M_3$, the dominant contribution for generating the lepton asymmetry stems from the decay of $N_1$ only, since the processes mediated by $N_1$ before its out-of-equilibrium decay, abolish the lepton asymmetry created by the decay of $N_2, N_3$ at $T \sim M_1$.
Therefore we have to solve two coupled Boltzmann equations involving $Y_{N_1}$ and $Y_{B-L}$. The simultaneous solution of the first and second Boltzmann equations yield comoving density $Y_{N_{1}}$ of the lightest heavy RH-neutrino $N_{1}$ and comoving density $Y_{B-L}$ of $B-L$ asymmetry respectively. Here $Y_{N_{1}}$  ($Y_{B-L}$) is defined as actual number density of $N_{1}$ (${B-L}$ asymmetry) divided by the entropy density $\bar{s}$ of the universe. Entropy density $\bar{s}$ can be written as : 
\bea
\bar{s} = \frac{2 \pi^2}{45} g_{eff} T^3 \,.
\eea

Here $T$ is the temperature \footnote{Not to be confused with aforementioned $T$-parameter.} and $g_{eff}$ is the total effective degrees of freedom which includes all the physical particles of the model. Detailed calculation of $g_{eff}$ is given in appendix \ref{app:D}.

 In general, the Boltzmann equations for $N_{1}$ and  the $(B-L)$ asymmetry can be written as \cite{Plumacher:1996kc},
 \bea
 \frac{\text{d} Y_{N_{1}}}{\text{d} z} &=& - \frac{z}{s \hspace{1mm} H(M_{1})} \Big[ \Big(\frac{Y_{N_{1}}}{Y^{eq}_{N_{1}}} - 1\Big)[\gamma_{D_{1}} + 2 \gamma^{1}_{\phi,s} + 4 \gamma^{1}_{\phi,t}]\Big] \,,
  \label{boltzmann_eq1}
 \eea
 \bea
 \frac{\text{d} Y_{B-L}}{\text{d} z}  &=& - \frac{z}{s \hspace{1mm} H(M_{1})} \Big[\left\{\frac{1}{2} \frac{Y_{B-L}}{Y^{eq}_{l}} + \epsilon ~\Big(\frac{Y_{N_{1}}}{Y^{eq}_{N_{1}}} - 1\Big) \right\} \gamma_{D_{1}}  \nonumber \\
 &&+ \frac{Y_{B-L}}{Y^{eq}_{l}}\left\{2 \gamma_{N,s} + 2 \gamma_{N,t} + 2 \gamma^{1}_{\phi,t} + \frac{Y_{N_{1}}}{Y^{eq}_{N_{1}}}  \gamma^{1}_{\phi,s} \right\} \Big] \,,
 \label{boltzmann_eq2}
 \eea
where $z = \frac{M_1}{T}$ and $H(M_1)$ is the Hubble parameter at $T = M_1$ : \\ $H(T=M_{1}) = 1.66 ~g_{eff}^{1/2} \frac{T^{2}}{M_{\rm{Pl}}}|_{T=M_{1}} $, $M_{\rm{Pl}} = 10^{19}$ GeV being Planck scale. $Y_{N_1}^{eq}, Y_l^{eq}$ are the comoving densities at equilibrium. We solve these two equations with initial conditions :
\bea
Y_{N_1}(0) = Y_{N_1}^{eq} , ~{\rm and}~ Y_{B-L}(0) = 0 \,.
\eea
at $T >> M_1$.

\begin{figure}[htpb!]{\centering
\subfigure[]{
\includegraphics[scale=0.1]{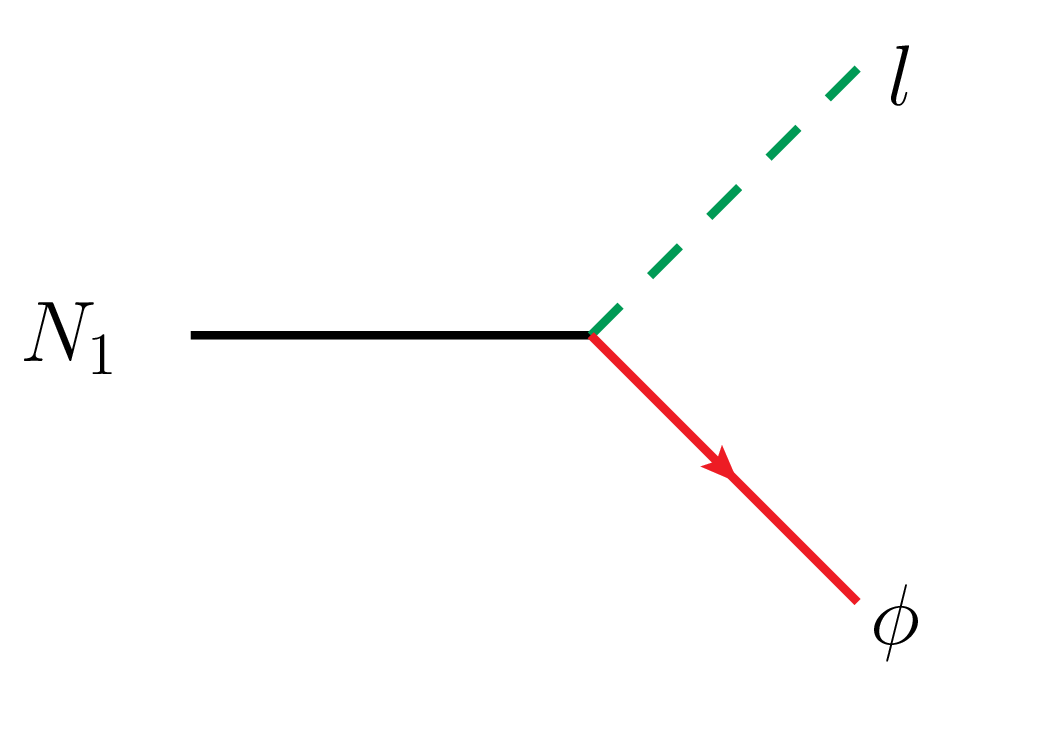}}
\subfigure[]{
\includegraphics[scale=0.1]{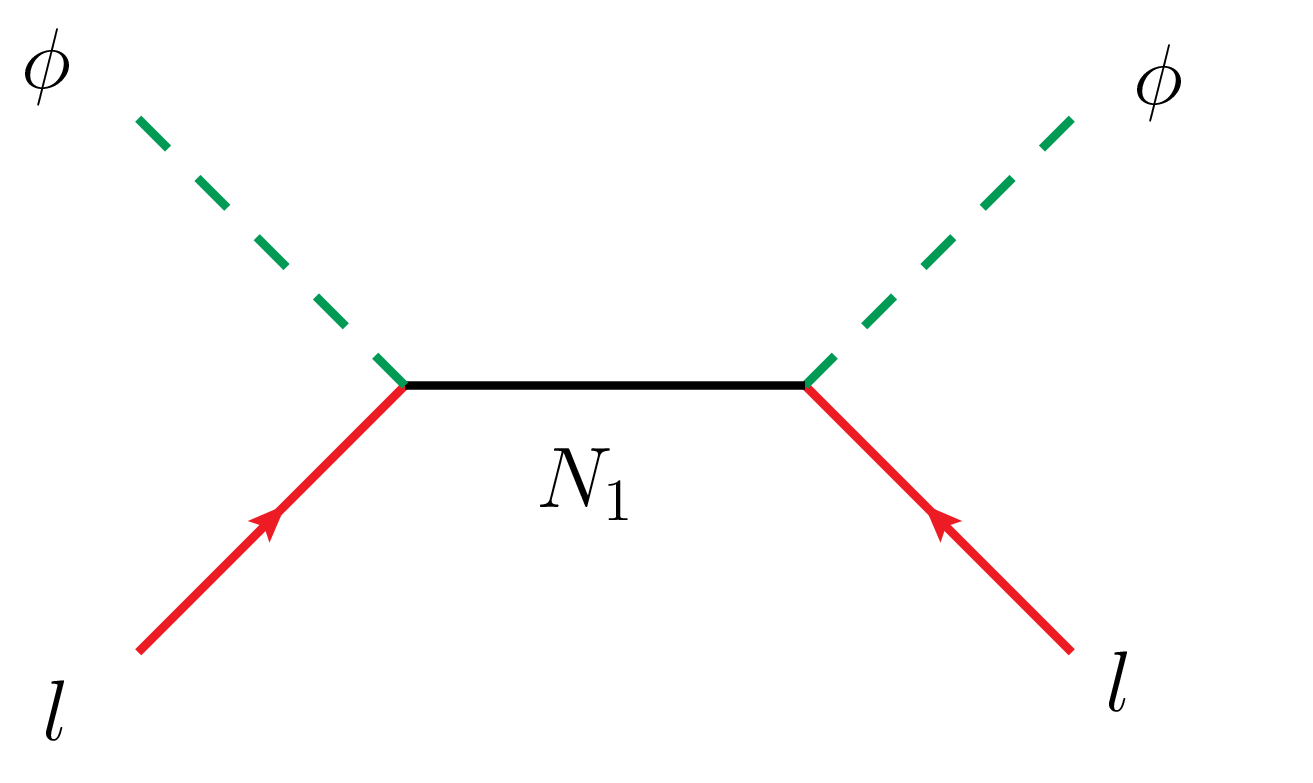}}
\subfigure[]{
\includegraphics[scale=0.1]{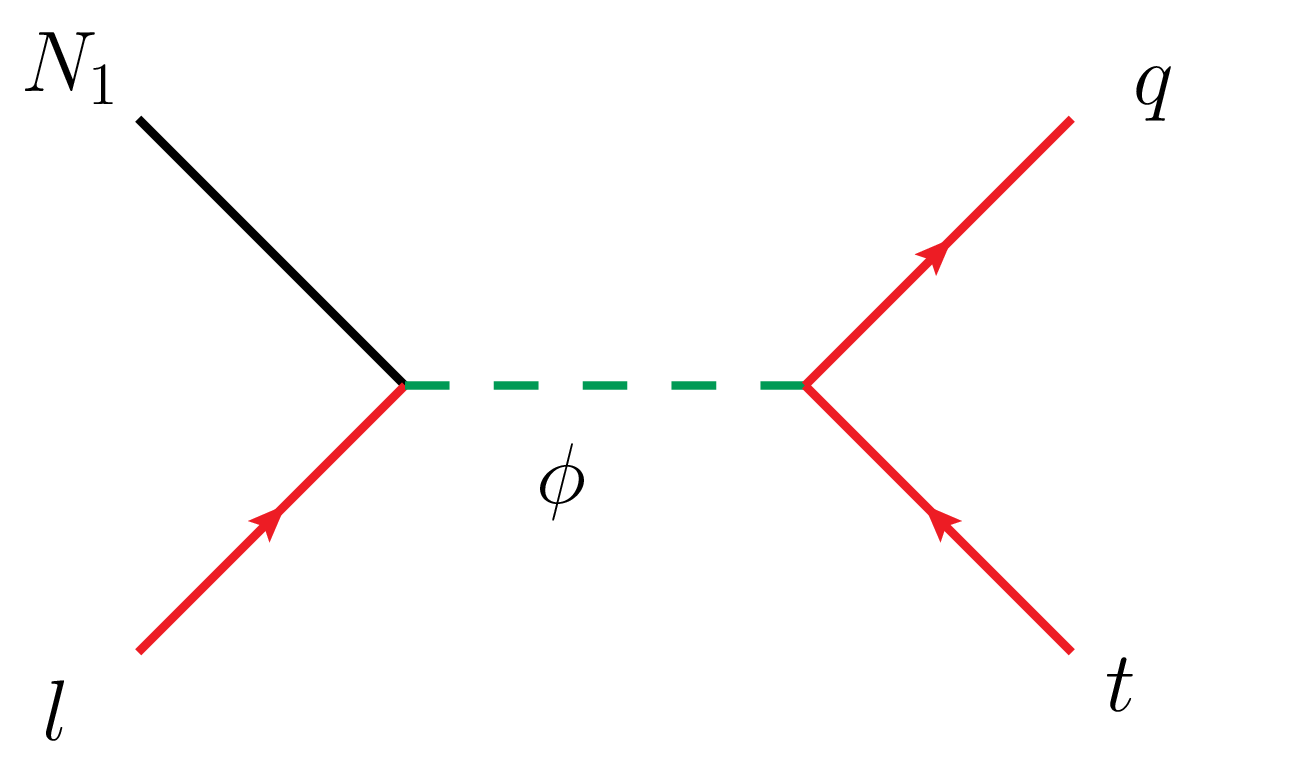}} \\
\subfigure[]{
\includegraphics[scale=0.1]{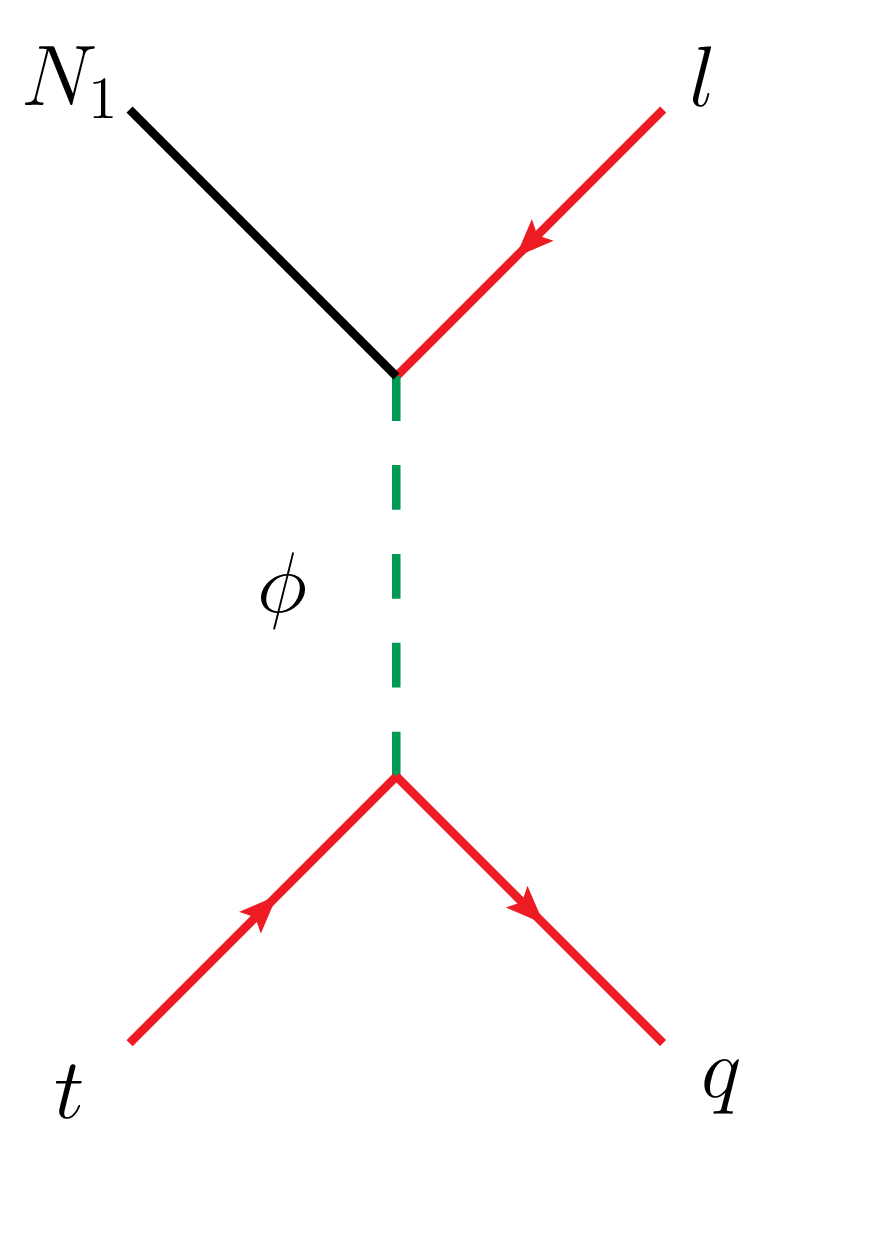}}
\subfigure[]{
\includegraphics[scale=0.1]{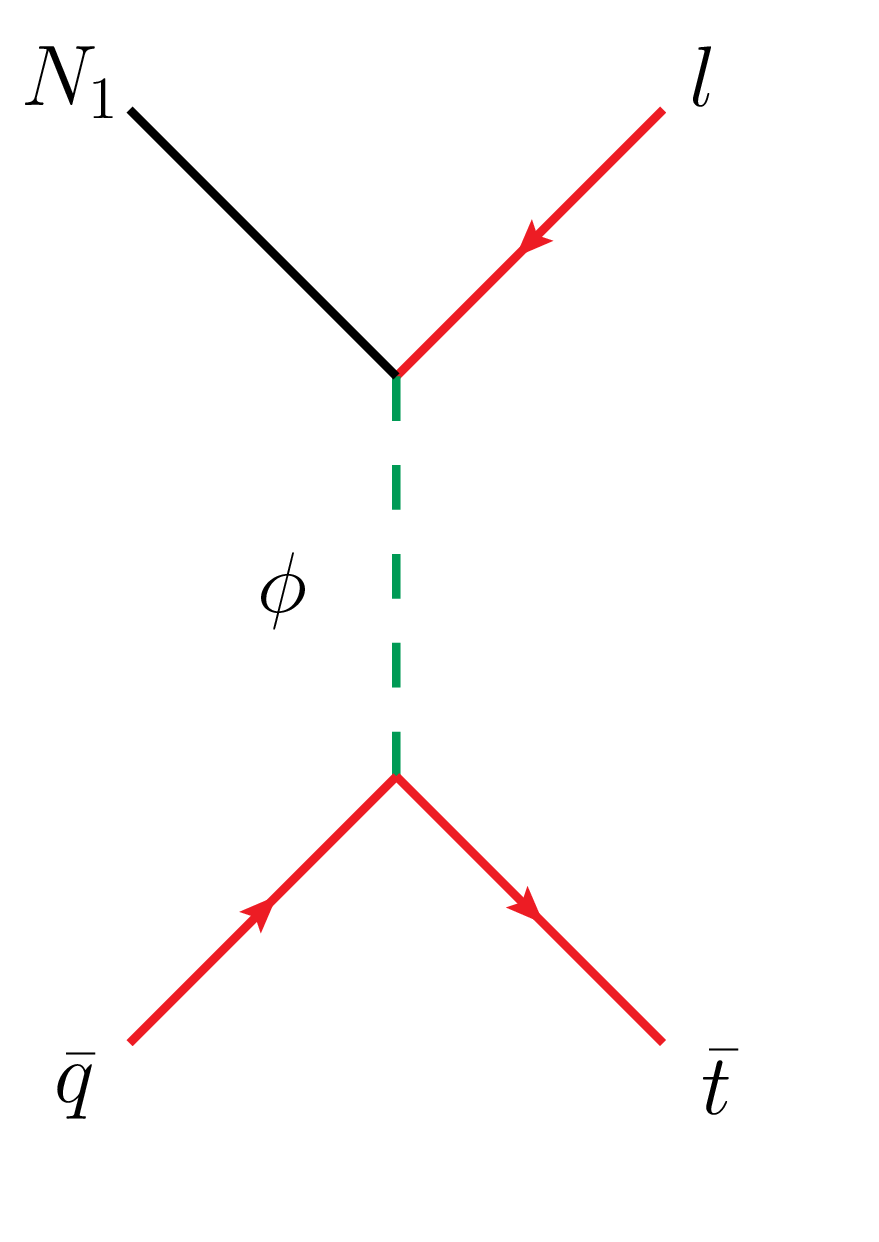}}
\subfigure[]{
\includegraphics[scale=0.1]{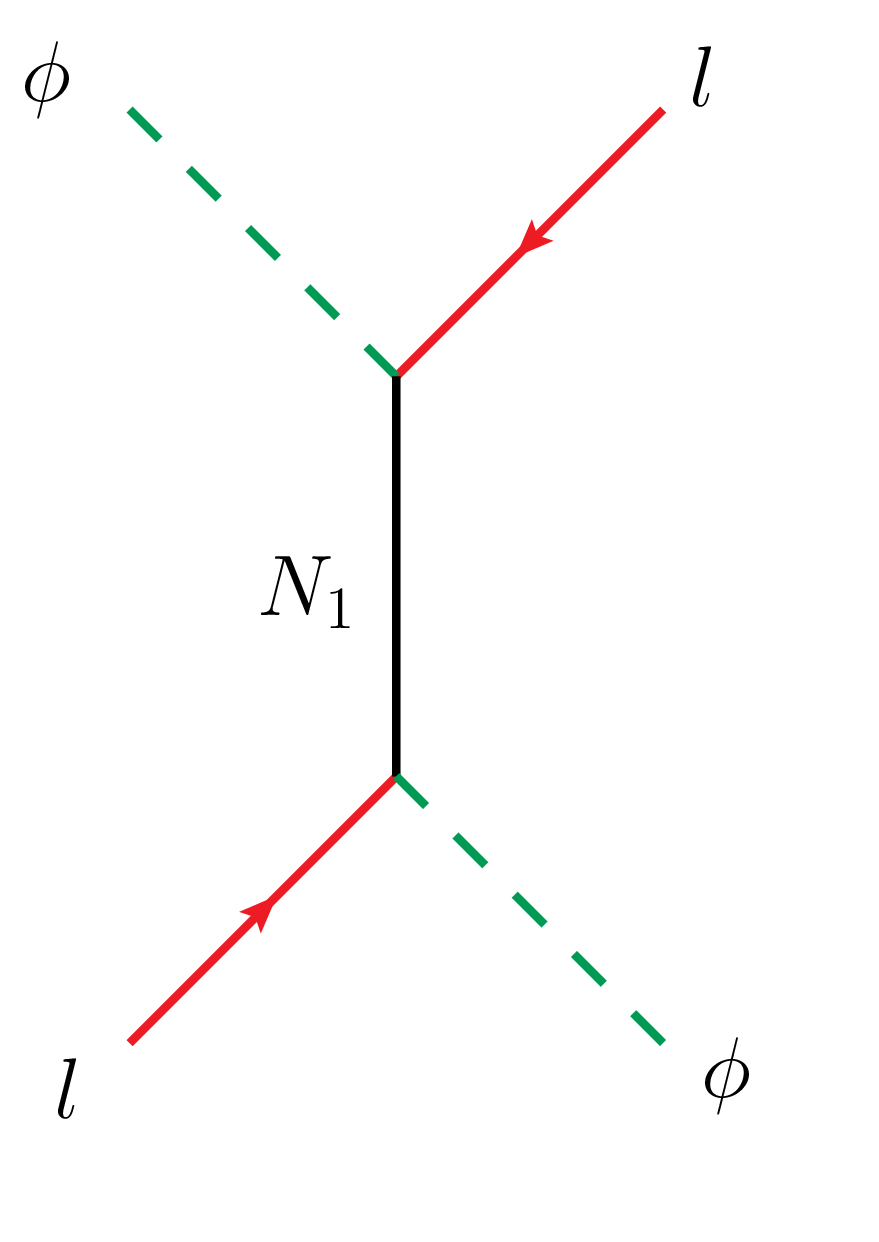}}} 
\caption{(a) Decay of lightest heavy RH-neutrino $N_{1}$ (contributes to $\gamma_{D_{1}}$), (b) $\Delta L = 2$, $s$-channel scattering via $N_{1}$ (contributes to $\gamma_{N,s}$), (c) $\Delta L = 1$, $s$-channel scattering via Higgs (contributes to $\gamma^{1}_{\phi,s}$), (d) and (e) $\Delta L = 1$, $t$-channel scattering via Higgs (contributes to $\gamma^{1}_{\phi,t}$), (f) $\Delta L = 2$, $t$-channel scattering via $N_{1}$ (contributes to $\gamma_{N,t}$). }
\label{diagram}
\end{figure}

Different $\gamma$s in eq.(\ref{boltzmann_eq1}) and eq.(\ref{boltzmann_eq2}) are space-time densities of the scattering processes at equilibrium depicted in fig.\ref{diagram}.
In the first Boltzmann equation (eq.(\ref{boltzmann_eq1})), $\gamma_{D_{1}}$ denotes the contribution from the decay of $N_{1}$. $\gamma^{1}_{\phi,s}$ and $\gamma^{1}_{\phi,t}$ originate from the lepton number-violating ($\Delta L = 1$) $s$-channel and $t$-channel washout processes via Higgs-mediation. The factor of $``2"$ in front of $\gamma^{1}_{\phi,s}$ comes due to the Majorana nature of $N_{1}$. The factor  $``4"$ in front of $\gamma^{1}_{\phi,t}$ accounts for the Majorana nature of $N_1$ as well as two $t$-channel washout scattering processes mediated by $N_{1}$ ($N_{1} t \rightarrow l q$ and $N_{1} \bar{q} \rightarrow l \bar{t}$) \cite{Buchmuller:2004nz}. $\gamma^{1}_{\phi,s}$ and $\gamma^{1}_{\phi,t}$ also contribute in the second Boltzmann equation. $\gamma_{N,s}$ and $\gamma_{N,t}$ in eq.(\ref{boltzmann_eq2}) come from $\Delta L = 2$ lepton number-violating $s$-channel  and $t$-channel scattering processes via $N_{1}$. The expressions of $\gamma_{D_{1}}$, $\gamma^{1}_{\phi,s}$, $\gamma^{1}_{\phi,t}$, $\gamma_{N,s}$ and $\gamma_{N,t}$ can be found in appendix \ref{app : B}.  However, in our model, since $\phi_3$ does not couple with the quarks from the requirement of zero FCNC (see the quantum number assignments in table \ref{tab:1}), no contributions will be drawn from  
$\gamma^{1}_{\phi,s}$ and $\gamma^{1}_{\phi,t}$ (fig.s 1(c), 1(d), 1(e)). Only surviving processes contributing to the washout will be $s$-channel and $t$-channel processes mediated by $N_1$ (fig.s 1(b), 1(f)). In our model, due to the quantum number assignment, $\phi \equiv \phi_3$ in fig.\ref{diagram}.

The number densities of particles with mass $M$ and temperature $T$ can be written as :
\bea
N_{eq} = \frac{g M^2 T}{2 \pi^2} K_2(\frac{M_1}{T}) \,.
\eea
$g$ being the number of degrees of freedom of corresponding particles, $K_2$ being second modified Bessel function of second kind.

\begin{figure}[htpb!]{\centering
\subfigure[]{
\includegraphics[scale=0.15]{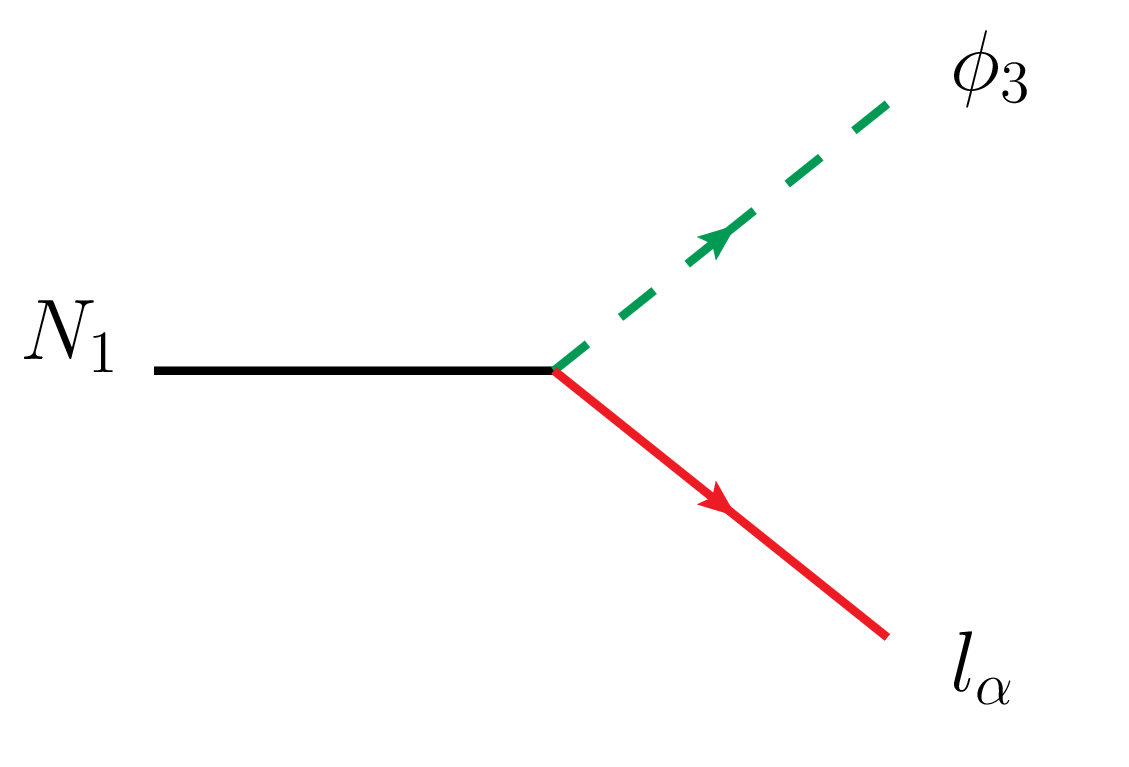}}
\subfigure[]{
\includegraphics[scale=0.15]{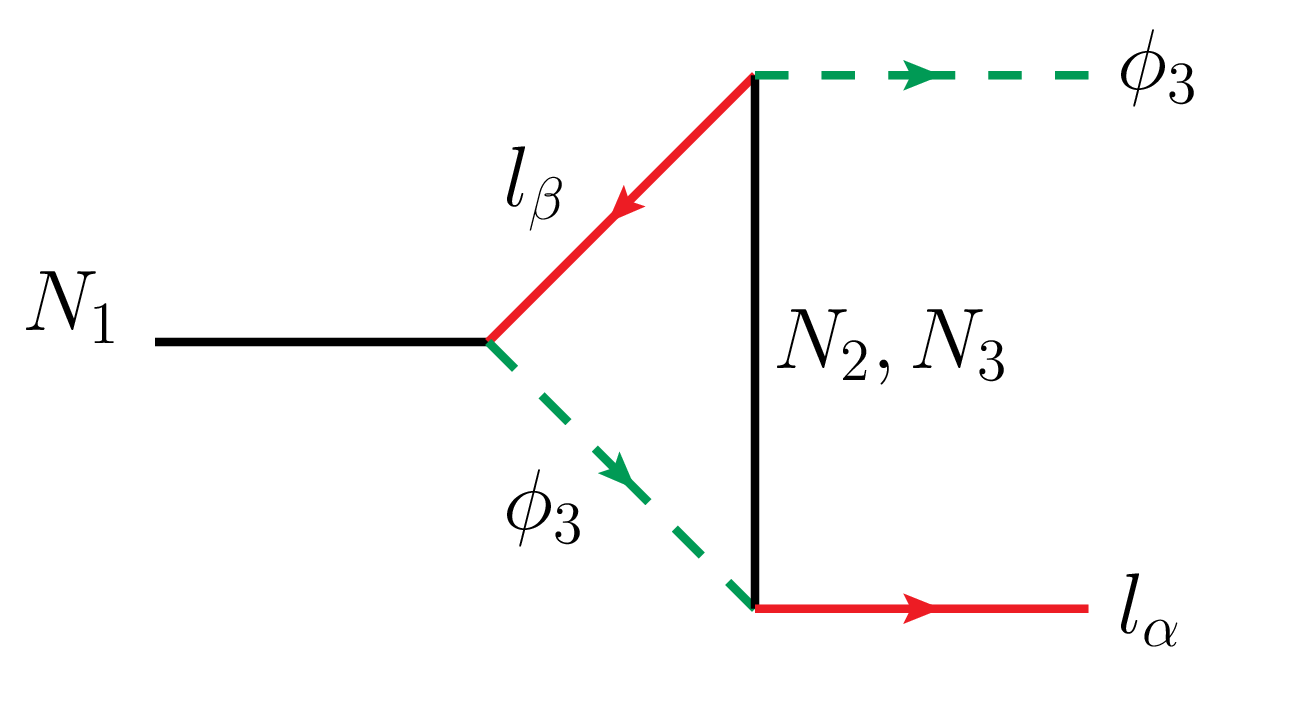}}
\subfigure[]{
\includegraphics[scale=0.15]{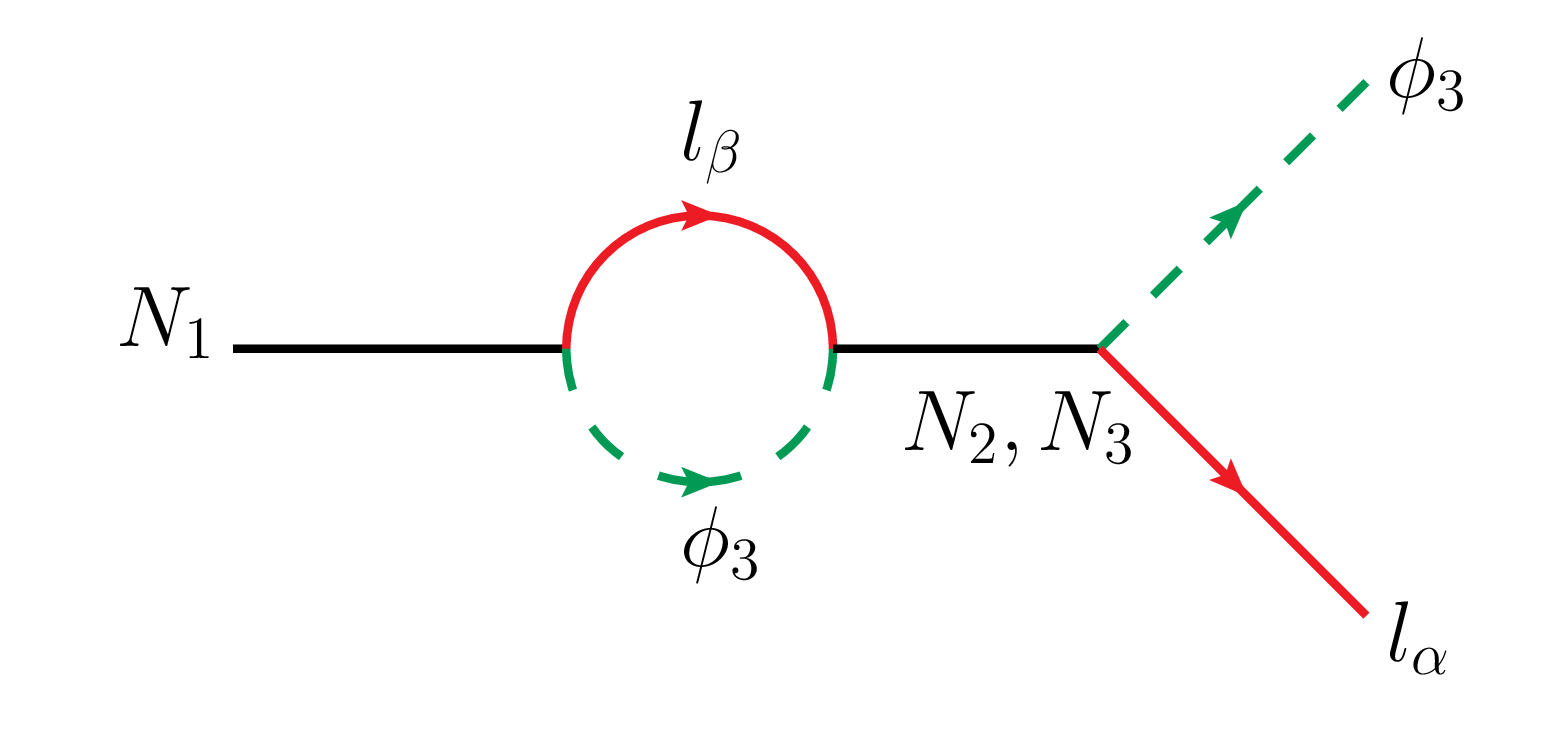}}} 
\caption{Diagrams contributing to the $CP$-asymmetry $\epsilon$ : (a) tree-level decay of $N_1$, (b) vertex correction, (c) self-energy diagram.}
\label{cp-asymmetry}
\end{figure}

$\epsilon_{\alpha \alpha}$ being the $CP$-asymmetry generated through the out-of-equilibrium decay of $N_{1}$ (decaying to $\phi_3$ and $l_\alpha$, $\alpha$ being the flavor of the lepton)) \cite{Davidson:2008bu}, the total lepton asymmetry can be computed by summing over the flavor indices, {\em i.e.} $\epsilon = \sum_\alpha \epsilon_{\alpha \alpha}$. Here the $CP$-asymmetry for a single flavor $\alpha$ can be calculated as \cite{Davidson:2008bu} :
\bea
\epsilon_{\alpha \alpha} := \frac{\Gamma(N_{1} \rightarrow \phi + l_{\alpha})  - \Gamma(N_{1} \rightarrow \bar{\phi} + \bar{l}_{\alpha})}{\Gamma(N_{1} \rightarrow \phi + l)  + \Gamma(N_{1} \rightarrow \bar{\phi} + \bar{l})}
\label{cp-asym}
\eea
 Considering the interference between the amplitudes of tree-level decay of $N_1$, (fig.2(a)), one-loop vertex-correction (fig.2(b)) and self-energy diagram (fig.2(c)), $\epsilon_{\alpha \alpha}$ can be calculated as \cite{Davidson:2008bu} :
\bea
\epsilon_{\alpha \alpha} &=& \frac{1}{8 \pi}~ \frac{1}{(Y^\dag Y)_{11}} \sum_j {\rm Im} \{Y^*_{\alpha 1} ~(Y^\dag Y)_{1j} ~Y_{\alpha j}\}~ g(x_j) \nonumber \\
&& + \frac{1}{8 \pi}~ \frac{1}{(Y^\dag Y)_{11}} \sum_j {\rm Im} \{Y^*_{\alpha 1} ~(Y^\dag Y)_{j1} ~Y_{\alpha j}\}~ \frac{1}{1-x_j} \,.
\label{cp_asym_1}
\eea
Here $\epsilon_{\alpha \alpha}$ in eq.(\ref{cp_asym_1}) can be written in this form assuming not too degenerate heavy neutrino spectrum, {\em i.e.} $M_i - M_j >> \Gamma_{N_1}$, $\Gamma_{N_1}$ being total tree-level decay width of $N_1$\footnote{Expression for $\Gamma_{N_1}$ can be found in appendix \ref{app : B}.}. The first term in eq.(\ref{cp_asym_1}) comes from the interference between the diagrams in fig.\ref{cp-asymmetry}, which violate both lepton flavor and lepton number. The second term in eq.(\ref{cp_asym_1}) violates lepton flavor, but conserves lepton number and hence does not contribute to the total lepton asymmetry $\epsilon$.

Thus total lepton asymmetry is computed by summing over the flavor indices \cite{Davidson:2008bu}:
\bea
\epsilon = \sum_\alpha \epsilon_{\alpha \alpha} = \frac{1}{8 \pi (Y^\dag Y)_{11}} \sum_j {\rm Im} \{\left[(Y^\dag Y)_{1j}\right]^2\} g(x_j) \,.
\eea
where $x_j = \frac{M_j^2}{M_1^2}$.

After taking the approximation $x >> 1$, 
\bea
g(x) = \sqrt{x} \left[\frac{1}{1-x} + 1 - (1+x)~ {\rm In} \left(\frac{1+x}{x}\right)\right] \rightarrow - \frac{3}{2 \sqrt{x}} \,. \nonumber \\
\eea
With $x_j = \frac{M_j^2}{M_1^2}$, $\epsilon$ becomes,
\bea
\epsilon = - \frac{3}{16 \pi (Y^\dag Y)_{11}} M_1 \left[ \frac{{\rm Im} \{(Y^\dag Y)_{12}^2\}}{M_2} + \frac{{\rm Im} \{(Y^\dag Y)_{13}^2\}}{M_3} \right] \,.
\label{lepton_asy}
\eea
 
Expressions for $(Y^\dag Y)_{11},~ {\rm Im} \{(Y^\dag Y)_{12}^2\},~ {\rm Im} \{(Y^\dag Y)_{13}^2\}$ are relegated to appendix \ref{app:C}.

The lepton asymmetry generated in the out-of-equilibrium decay of $N_1$, is converted into baryon asymmetry through $(B+L)$ violating sphelaron transitions \cite{Klinkhamer:1984di,Kuzmin:1985mm}. The conversion of lepton asymmetry to baryon asymmetry being terminated at the freeze-out temperature of the sphelaron process, $T_{\rm sph} \sim 150$ GeV \cite{Burnier:2005hp}, the resultant baryon number is computed at $T_{\rm sph}$ as \cite{Khlebnikov:1988sr}:  
\bea
Y_{B} = \bigg( \frac{8 N_{f} + 4 N_{H}}{22 N_{f} + 13 N_{H} } \bigg) Y_{B-L} (z_{\rm sph}).
\label{baryon-asym-con}
\eea
where $N_{f}$ is the number of generations of the fermion families and $N_{H}$ is the number of the Higgs doublets and $Y_{B-L} (z_{\rm sph})$ is the solution of Boltzman equations at $z = z_{\rm sph} = \frac{M_1}{T_{\rm sph}}$. For our model $N_f = 3 , N_H = 3$.


\section{Analysis and Results}
\label{sec:5}
To analyse the multi-dimensional parameter space compatible with the aforementioned theoretical and experimental constraints, we have considered the model parameters $\tan\beta , \gamma, \gamma_1 , \gamma_2, \alpha_3,  M_h,  M_{H_1}, M_{H_2}, M_{A_1}, M_{A_2}, M_{H_1^{\pm}}, M_{H_2^{\pm}}$ as independent, and have varied them (except $M_h$ and $\alpha_3$) within the following window : 
\bea
&& 2.5 < \tan\beta < 10.0 , ~  - \pi < \gamma, \gamma_1 , \gamma_2 < \pi,  \nonumber \\
&& 300~ {\rm GeV} < M_{H_1}, M_{H_2}, M_{A_1}, M_{A_2}, M_{H_1^{\pm}}, 
 M_{H_2^{\pm}} < 500~ {\rm GeV} \,.
\eea
The dependent parameters $\alpha_1, \alpha_2$ can be expressed in terms of the independent ones. Mixing angles in the $CP$-even sector $\alpha_1, \alpha_2 $ are fixed by the relations $\alpha_1 = \gamma , \alpha_2 + \beta = \frac{\pi}{2}$ at the {\em Alignment limit} and we have fixed $M_h = 125$ GeV and $\alpha_3 = \frac{\pi}{4}$ throughout the analysis.
\begin{figure}[htpb!]{\centering
  \subfigure[]{
 \includegraphics[width=7.5cm,height=7.5cm, angle=0]{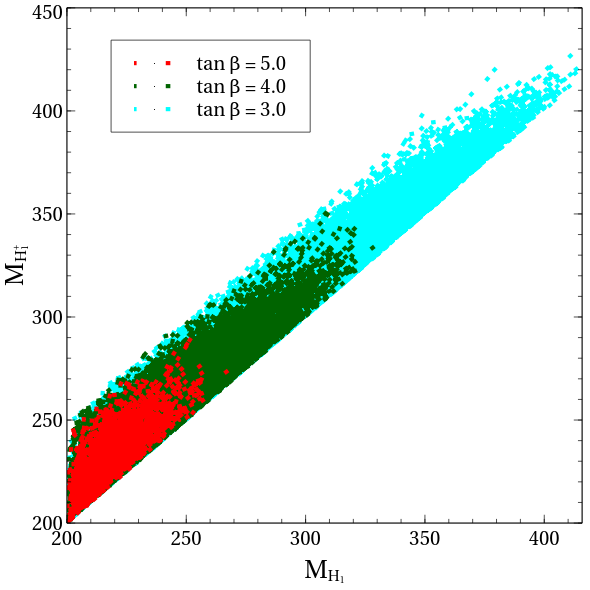}}
  \subfigure[]{
 \includegraphics[width=7.5cm,height=7.5cm, angle=0]{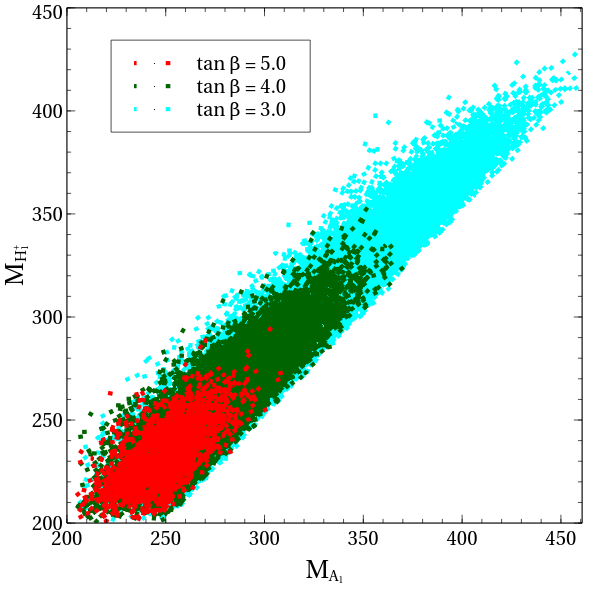}} \\
  \subfigure[]{
 \includegraphics[width=7.5cm,height=7.5cm, angle=0]{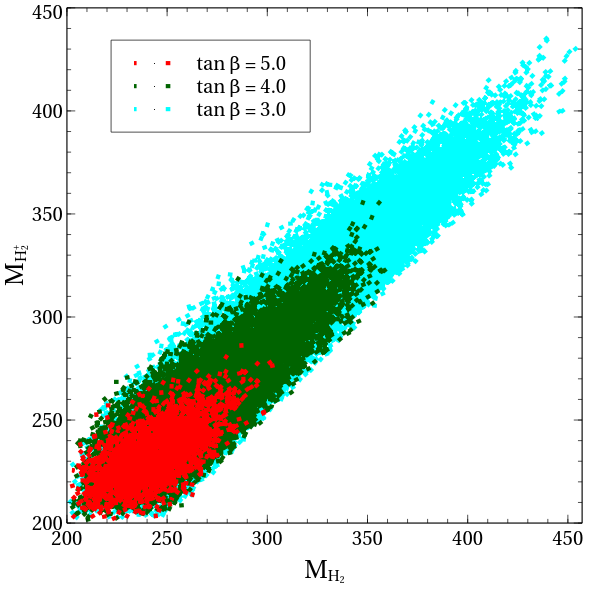}}
  \subfigure[]{
 \includegraphics[width=7.5cm,height=7.5cm, angle=0]{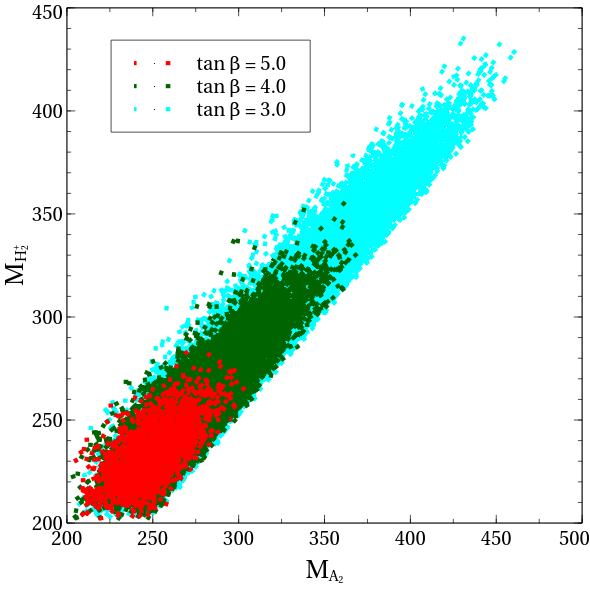}}}
 \caption{This figure depicts parameter space in $M_{H_1^+}~ {\rm vs.} ~M_{H_1}$ plane (upper panel left), $M_{H_1^+} ~{\rm vs.}~M_{A_1}$ plane (upper panel right), $M_{H_2^+}~ {\rm vs.}~ M_{H_2}$ plane (lower panel left), $M_{H_2^+}~ {\rm vs.}~ M_{A_2}$ plane (lower panel right). The cyan, green and red regions represent the parameter space allowed by all constraints in the scalar sector (mentioned in sec. \ref{sec:2}) for $\tan \beta = 3,~4,~5$ respectively.}
 \label{fig1}
 \end{figure}
After doing an extensive scan over the parameter space, subject to all the aforementioned constraints listed in sec.\ref{sec:2}, larger values of $\tan \beta$, {\em i.e.} $\tan \beta > 5.5$, are ruled out particularly from the perturbativity constraints ($|\lambda_i| \leq 4 \pi$) on the quartic couplings. Therefore we have presented the results for three discrete values of $\tan \beta$, {\em i.e.} $\tan \beta = 3,~4,~5$. Since doublet $\phi_3$ is responsible for generating masses of the light neutrinos due to $Z_3$-symmetry (eq.(\ref{MDirac})), $\tan \beta$ plays a crucial role and subsequently enters into the calculation of lepton asymmetry (eq.(\ref{lepton_asy})). Throughout the analysis, we have used those values of $\tan \beta$, which are filtered out by the constraints in the scalar sector mentioned in sec. \ref{sec:2}. 
 At exact {\em Alignment limit}, $h$ being the SM-like Higgs boson, masses of other non-standard heavier scalars range from 200-450 GeV depending on $\tan \beta$. Fig.\ref{fig1} depicts the parameter space spanned by one charged and one neutral scalar(s). Lower the value of $\tan \beta$, parameter space with higher masses of non-standard scalars becomes accessible. The masses of non-standard scalars are functions of the quartic couplings $\lambda_i$s ($i=1,2,...12$) and the mixing angles $\beta, \gamma, \gamma_1, \gamma_2, \alpha_1, \alpha_2, \alpha_3$ etc. Other mixing angles other than "$\beta$" also play a role in constructing the physical masses. Thus a simple dependence of the physical masses on $\tan \beta$ alone cannot be derived by neglecting the effect of varying other parameters. Here one can at most comment that the parameter space with lower values of $\tan \beta$ and higher values of physical masses comply with all the constraints mentioned in section III.  It is clearly evident that with rise of $\tan \beta$, the available parameter space consistent with all the constraints, shrinks from the cyan colored region with $\tan \beta = 3$ to red colored region with $\tan \beta = 5$ in fig.\ref{fig1}. Constraints coming from $T$-parameter restrict the mass-splittings between the heavy neutral and charged scalars within 50 GeV. These mass splittings result in the sharp edges in the plots of fig.\ref{fig1}.

After putting an upper bound on $\tan \beta$ from the constraints in the scalar sector, let us now move on to explore the status of the parameter space in the neutrino sector. As discussed earlier, using CI parametrization the real and imaginary parts of Yukawa couplings are solved in terms of $v_3$, the elements in the PMNS matrix, complex angles $\theta, \phi, \psi$ and $M_1, M_2, M_3$. The elements in the $U_{\rm PMNS}$ matrix, constrained by neutrino oscillation data, are fixed at their central values \cite{Esteban2019},
\bea
&&\sin^2 \theta_{12} = 0.31 , ~ \sin^2 \theta_{23} = 0.58 , ~ \sin^2 \theta_{13} = 0.02241, \nonumber \\
&&\Delta m_{21}^2 = 7.39 \times 10^{-5} {\rm eV}^2 , ~ \Delta m_{31}^2 = 2.525 \times 10^{-3} {\rm eV}^2 , ~ \delta_{\rm CP} = 215^\circ \,.
\eea
Assuming the phases associated with the three complex angles $\theta, \phi, \psi$ in the matrix $\mathcal{O}$ (eq.(\ref{ortho_mat})) to be zero, they are varied within the region : $- \pi < \phi < \pi, -\pi < \psi < \pi $ and we fix $\theta$ at $\theta = \frac{\pi}{4}$ for simplicity. The variation of these angles will in turn incorporate variations in the real ($y_{jR}$) and imaginary parts ($y_{jI}$) of Yukawa couplings in $M_D$, which are absolutely compatible with the neutrino oscillation data \cite{Esteban2019}. 

\begin{figure}[!htpb]
\includegraphics[scale=0.6]{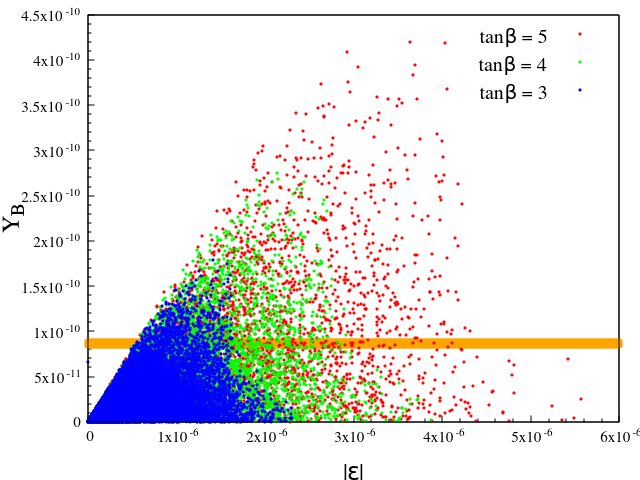}
\caption{Parameter space in $Y_B-|\epsilon|$ plane with varying $M_1, M_2, M_3$. Blue, green and red points correspond to the regions with $\tan \beta = 3,~4,~5$. The Orange band signifies $2 \sigma$-deviation from the central value of $Y_B$ mentioned in introduction.}
\label{fig3}
\end{figure}

In thermal leptogenesis there exists a lower bound of $10^9$ GeV on $M_1$, {\em i.e.} $M_1 \geq 10^9$ GeV, which is known as {\em Davidson-Ibarra bound} \cite{Davidson:2002qv, Davidson:2008bu}. In addition, we consider the heavy neutrino masses to be hierarchical. In fig.\ref{fig3}, the red, green and blue points in "$Y_B-|\epsilon|$" plane represent the points filtered out by neutrino oscillation data for $\tan \beta = 5, ~4, ~3$ respectively (with $10^9 ~{\rm GeV} < M_1 < 10^{11} ~{\rm GeV}, 10^{13} ~{\rm GeV} < M_2, M_3 < 10^{16}~ {\rm GeV}$). The narrow orange band represents the region of the parameter space consistent with the current baryon asymmetry data. The red, green and blue points lying within the orange band thus comply with the neutrino oscillation data and observed baryon  asymmetry of the Universe. Therefore the red, blue and green points above the orange band are ruled out by the current baryon asymmetry data. Whereas for points lying below the orange band, leptogenesis fails to produce adequate matter anti-matter asymmetry. Due to the interplay of the model parameters like real and imaginary parts of Yukawa couplings, $\tan \beta, ~ M_1, ~ M_2, ~ M_3$, the allowed parameter space gets larger for higher values of $\tan \beta$. Fig.\ref{fig3} shows that with increase of $\tan \beta$, larger values of $|\epsilon|$ are attainable.  The common parameter space in $Y_B$ vs. $|\epsilon|$ plane in fig.\ref{fig3} (crowded by the blue, green and red points lying within the orange band), indicates that the production of sufficient baryon asymmetry requires $|\epsilon| \sim 10^{-6}$ for $\tan \beta = 3,4,5$. \footnote{Here we have not considered the region with $|\epsilon| > 1.5 \times 10^{-6}$  (crowded only with green or red points) in fig.\ref{fig3}, since we aim to explore the parameter space common for all values of $\tan \beta$.}  While its order of magnitude remains the same for all three values of tan$\beta$ (= 3,4,5) taken, the most restrictive bound $1 \times 10^{-6} < |\epsilon| < 1.5 \times 10^{-6}$ is obtained for tan$\beta$ = 3 (blue points in fig.\ref{fig3}). In other words, the points lying inside the orange band and obeying $1 \times 10^{-6} < |\epsilon| < 1.5 \times 10^{-6}$ lead to the requisite 
$Y_B$ irrespective of tan$\beta$. 
\begin{figure}[htpb!]{\centering
  \subfigure[]{
 \includegraphics[width=8cm,height=7cm, angle=0]{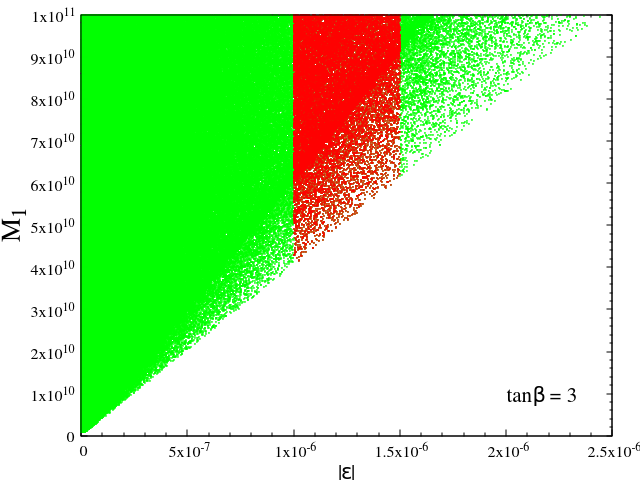}}
  \subfigure[]{
 \includegraphics[width=8cm,height=7cm, angle=0]{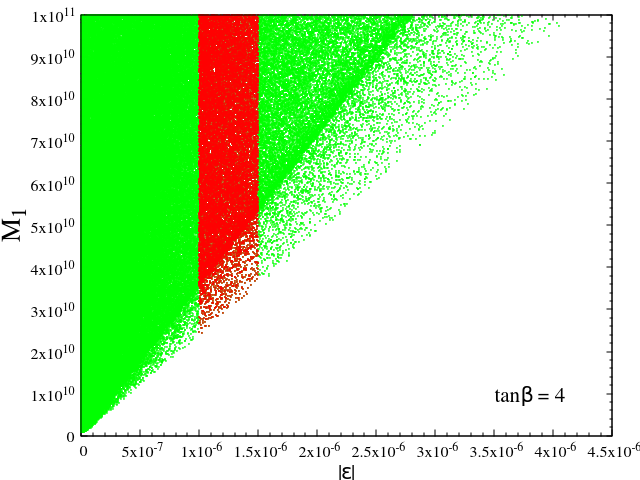}}
  \subfigure[]{
 \includegraphics[width=8cm,height=7cm, angle=0]{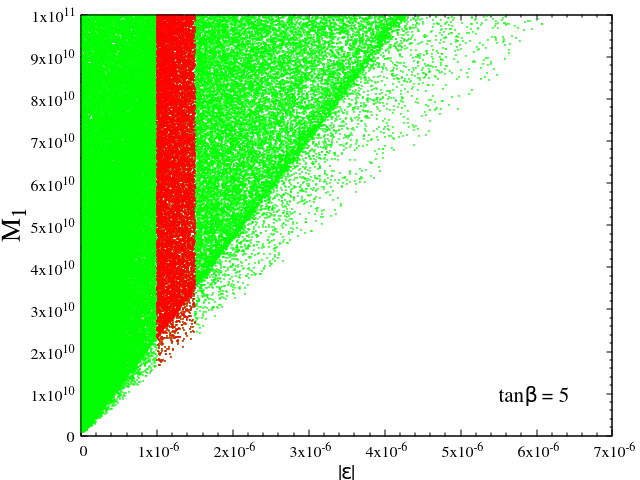}}}
 \caption{Region allowed by neutrino oscillation data in $M_1~ {\rm vs.} ~ |\epsilon|$ plane. Green region corresponds to the points with no constraint on $|\epsilon|$, whereas red points satisfy : $1 \times 10^{-6} < |\epsilon| < 1.5 \times 10^{-6}$. Figures are drawn for $\tan \beta = 3$ (upper panel left), $\tan \beta = 4$ (upper panel right), $\tan \beta = 5$ (lower panel).}
 \label{fig2}
 \end{figure}

In fig.\ref{fig2}, the variation of $M_1$ with respect to $|\epsilon|$ have been shown for three different values of $\tan \beta$. For the green points, only neutrino oscillation data has been satisfied. The red band on top of the green region signifies the reduced parameter space in $M_1~ {\rm vs. } ~ |\epsilon|$ plane after applying the aforementioned bound on $|\epsilon|$. Here it is needed to be clarified that all the red points do not lead to 100 $\%$ of the observed baryon asymmetry. Some points in the red region, depending on the other parameters in the Boltzman equation, indeed lead to 100$\%$ of the baryon asymmetry. The green region at the left of the red band corresponds to the under production of baryon asymmetry, since from fig.\ref{fig3}, for $|\epsilon| < 1 \times 10^{-6}$, there is hardly any point leading to exact (within the orange band) or excess (points above the orange band) baryon asymmetry. The green region at the right of the red band in fig.\ref{fig2}, partially corresponds to the under production of the same following fig.\ref{fig3}. Apart from this, another major fraction of the green region at the right of the red band refers to the overproduction of baryon asymmetry. A very small fraction of the green points at the right of the red band, corresponds to 100$\%$ baryon asymmetry. For higher $\tan \beta$, most of the green region at the right  contains points with overproduced baryon asymmetry (as can be seen from fig.\ref{fig3} also). Increase in $\tan \beta$ makes lower values of $M_1$ allowed, which are compatible with both neutrino oscillation data and the bound from lepton asymmetry. It can be inferred that this constraint on $|\epsilon|$ disfavors the portion of the parameter space with $M_1 < 4 \times 10^{10}$ GeV, $ 2.5 \times 10^{10}$ GeV, $ 1.5 \times 10^{10}$ GeV for $\tan \beta = 3,~4,~5$ respectively in our model. Thus the lower bound of $10^9$ GeV on $M_1$ for thermal leptogenesis is uplifted after being filtered out by all constraints. 

From eq.(\ref{lepton_asy}) it can be seen that for fixed values of Yukawa couplings and $M_1$, $|\epsilon|$ decreases with increasing $M_2$ and $M_3$. From fig.\ref{fig3} it can be concluded that the parameter space with $|\epsilon|$ smaller than $\sim 10^{-6}$ (and hence too large $M_2, M_3$ for fixed values of Yukawa couplings and $M_1$) cannot produce sufficient baryon asymmetry. Thus we choose to explore the parameter region of Yukawa couplings corresponding to $M_1 = 10^{11} $ GeV, and varied $M_2, M_3$ from $10^{13}$ GeV to $10^{16}$ GeV for all values of $\tan \beta$, so that adequate  baryon asymmetry can be produced. Real and imaginary parts of the Yukawa couplings $y_j$, already being compatible with neutrino oscillation data, get  additional constraints coming from the lepton asymmetry $|\epsilon|$ (eq.(\ref{lepton_asy})). All points satisfying the neutrino oscillation data and the constraints coming from lepton asymmetry, are further validated by the baryon asymmetry constraint (eq.(\ref{Bar_asy})). At this point, a clear distinction between the two constraints coming from the lepton asymmetry and observed baryon asymmetry is required. The parameter space compatible with the observed baryon asymmetry (orange band) in fig.(4) is a subset of the parameter space satisfied by the constraint on lepton asymmetry (includes points residing within, above and below the orange band). After satisfying the aforementioned bound on $|\epsilon|$, the baryon asymmetry is under produced and over produced for the green, blue and red points lying below and above the orange band respectively. Thus all points satisfying the lepton asymmetry bound may not comply with the observed baryon asymmetry data. This is consistent with eq.(\ref{boltzmann_eq2}), since the solution of the second Boltzman equation depends not only on $\epsilon$, but also on some other parameters like $\gamma_{N,s}, \gamma_{N,t}$ etc.

\begin{figure}[htpb!]{\centering
  \subfigure[]{
 \includegraphics[width=7.6cm,height=7cm, angle=0]{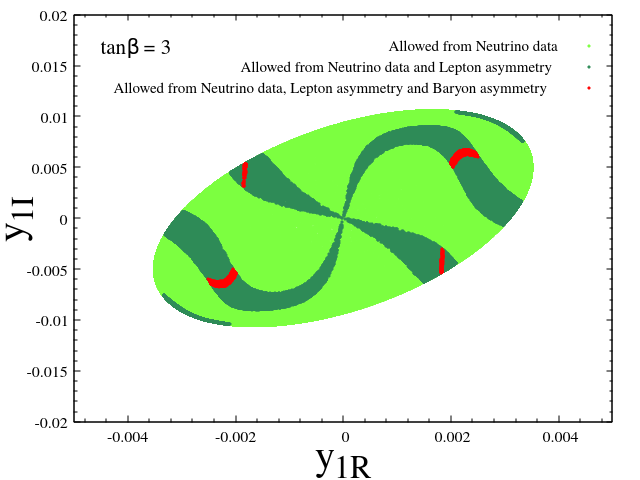}} 
   \subfigure[]{
 \includegraphics[width=8cm,height=7.6cm, angle=0]{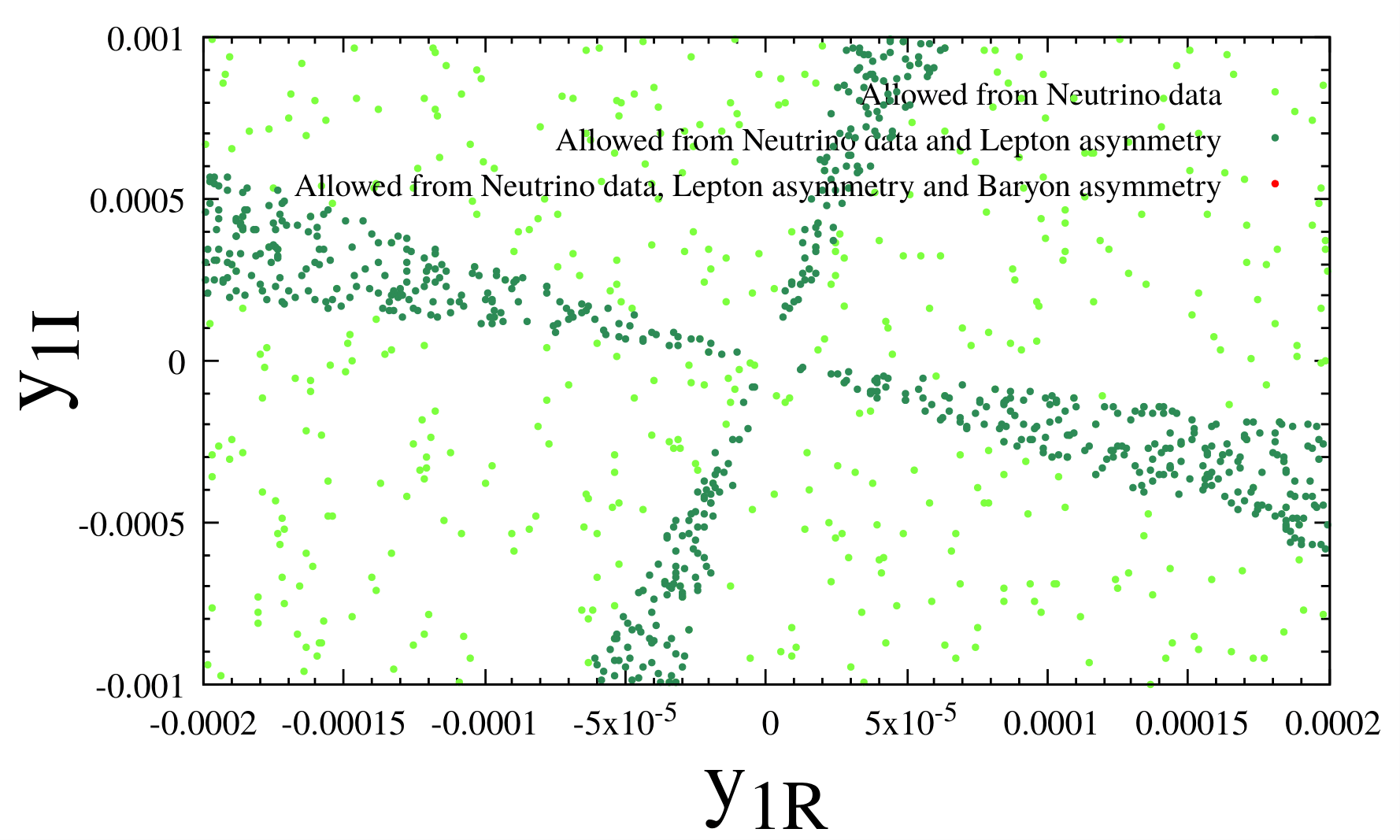}} \\
  \subfigure[]{
 \includegraphics[width=7.6cm,height=7cm, angle=0]{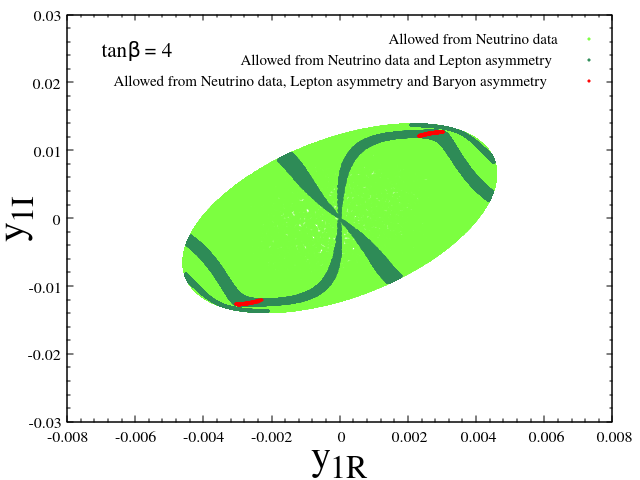}} 
  \subfigure[]{
 \includegraphics[width=7.6cm,height=7cm, angle=0]{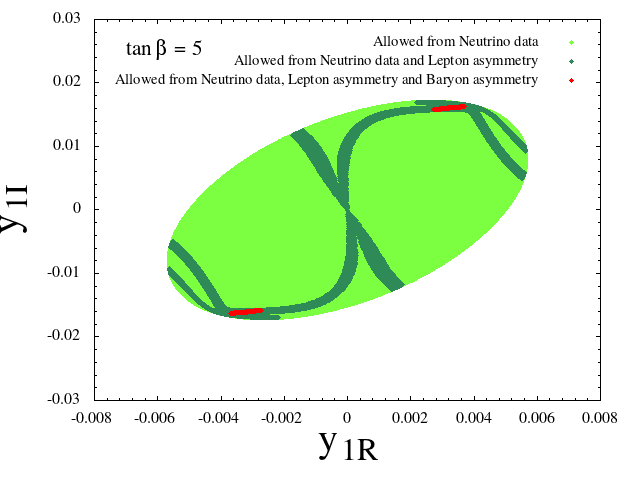}}} 
 \caption{Parameter space in real vs. imaginary part of Yukawa coupling $y_1$. Plots are given for $\tan \beta = 3$ (upper panel left), $\tan \beta = 4$ (lower panel left), $\tan \beta = 5$ (lower panel right). Light green, deep green and red region correspond to the parameter space allowed by only neutrino oscillation data, neutrino oscillation data + lepton asymmetry, neutrino oscillation data + lepton asymmetry + baryon asymmetry respectively. The right plot in the upper panel is the zoomed version of the figure at the left for $\tan \beta = 3$.}
 \label{fig4}
 \end{figure}
\begin{figure}[htpb!]{\centering
  \subfigure[]{
 \includegraphics[width=8cm,height=7cm, angle=0]{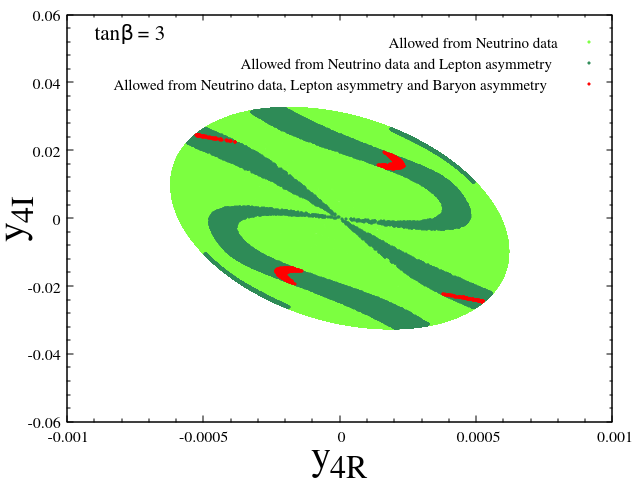}} 
  \subfigure[]{
 \includegraphics[width=8cm,height=7cm, angle=0]{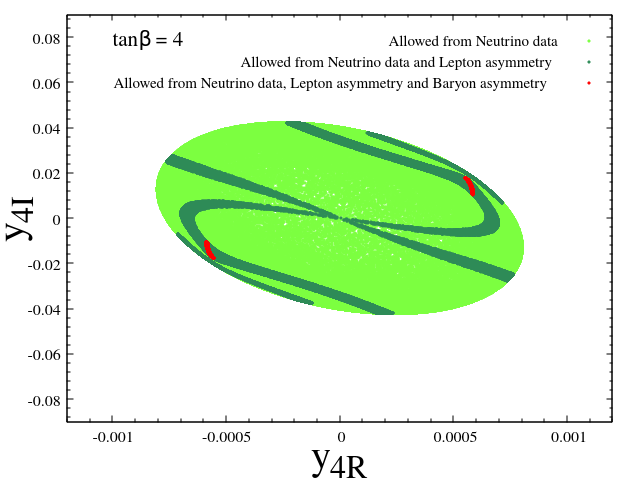}} \\
  \subfigure[]{
 \includegraphics[width=8cm,height=7cm, angle=0]{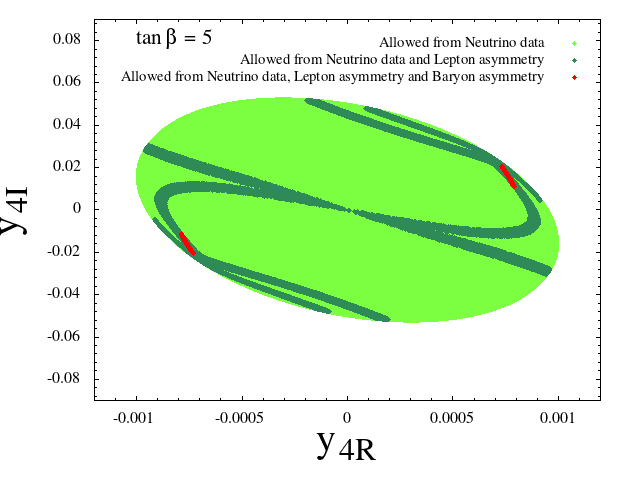}}} 
\caption{Parameter space in real vs. imaginary part of Yukawa coupling $y_4$. Plots are given for $\tan \beta = 3$ (upper panel left), $\tan \beta = 4$ (upper panel right), $\tan \beta = 5$ (lower panel). Light green, deep green and red region correspond to the parameter space allowed by only neutrino oscillation data, neutrino oscillation data + lepton asymmetry, neutrino oscillation data + lepton asymmetry + baryon asymmetry respectively. }
 \label{fig5}
 \end{figure}
\begin{figure}[htpb!]{\centering
  \subfigure[]{
 \includegraphics[width=8cm,height=7cm, angle=0]{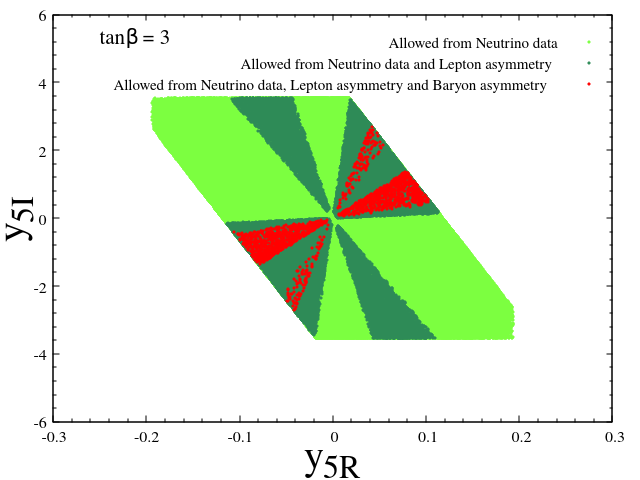}} 
  \subfigure[]{
 \includegraphics[width=8cm,height=7cm, angle=0]{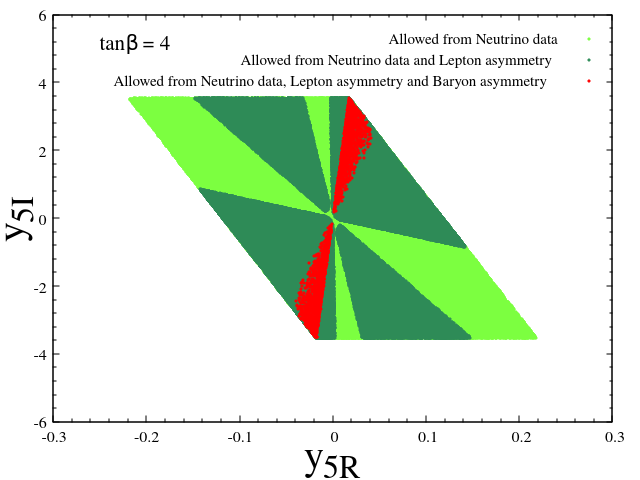}} \\
  \subfigure[]{
 \includegraphics[width=8cm,height=7cm, angle=0]{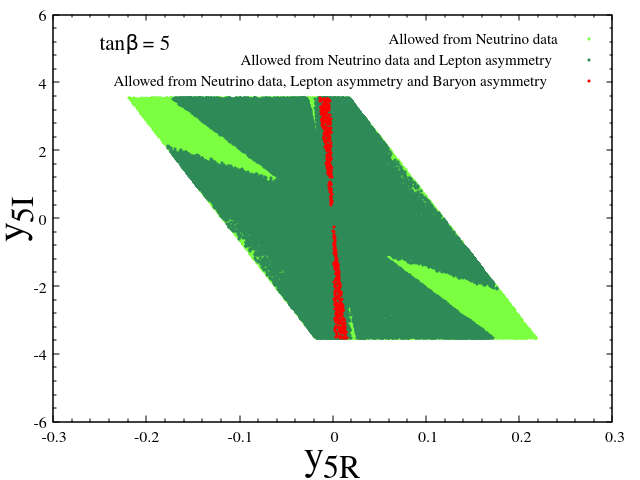}}} 
\caption{Parameter space in real vs. imaginary part of Yukawa coupling $y_5$. Plots are given for $\tan \beta = 3$ (upper panel left), $\tan \beta = 4$ (upper panel right), $\tan \beta = 5$ (lower panel). Light green, deep green and red region correspond to the parameter space allowed by only neutrino oscillation data, neutrino oscillation data + lepton asymmetry, neutrino oscillation data + lepton asymmetry + baryon asymmetry respectively. }
 \label{fig6}
 \end{figure}
\begin{figure}[htpb!]{\centering
  \subfigure[]{
 \includegraphics[width=8cm,height=7cm, angle=0]{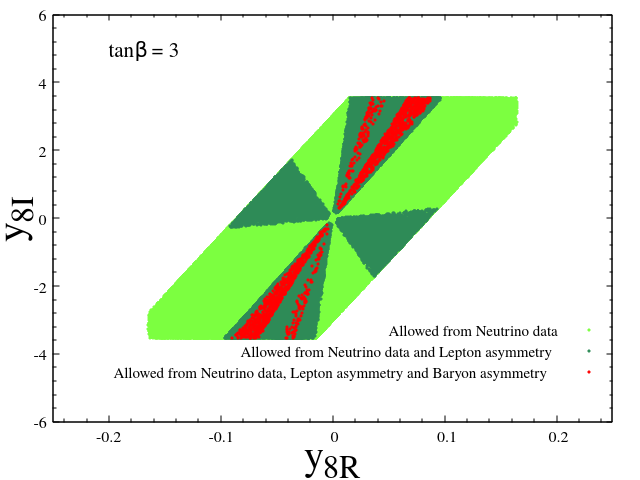}} 
  \subfigure[]{
 \includegraphics[width=8cm,height=7cm, angle=0]{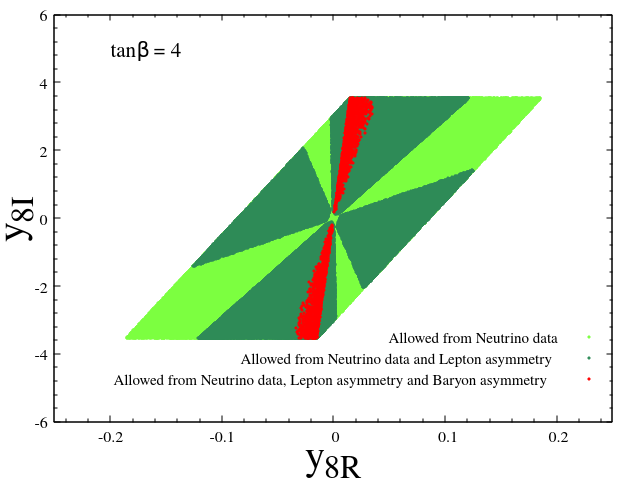}} \\
  \subfigure[]{
 \includegraphics[width=8cm,height=7cm, angle=0]{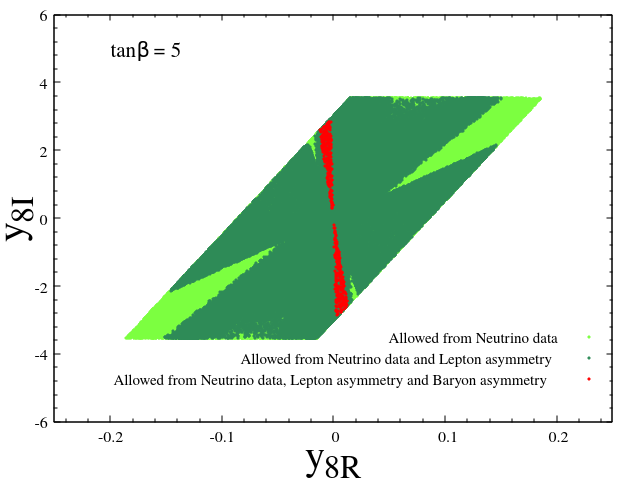}}} 
\caption{Parameter space in real vs. imaginary part of Yukawa coupling $y_8$. Plots are given for $\tan \beta = 3$ (upper panel left), $\tan \beta = 4$ (upper panel right), $\tan \beta = 5$ (lower panel). Light green, deep green and red region correspond to the parameter space allowed by only neutrino oscillation data, neutrino oscillation data + lepton asymmetry, neutrino oscillation data + lepton asymmetry + baryon asymmetry respectively. }
 \label{fig7}
 \end{figure}

From fig.\ref{fig4}, fig.\ref{fig5}, fig.\ref{fig6}, fig.\ref{fig7}, it can be inferred that the available parameter spaces spanned by the real and imaginary parts of the Yukawa couplings ($y_{j}$), at $\tan \beta =3,~4,~5$, shrink gradually after applying the following constraints sequentially : (i) neutrino oscillation data (light green region), (ii) lepton asymmetry (deep green region), (iii) baryon asymmetry (red region). Thus the bound on the Yukawa couplings coming from baryon asymmetry comes out to be the most stringent among all. From eq.(\ref{lepton_asy}) and eq.(\ref{epsi_part}) it can be computed that for a given set of positive Yukawa couplings, $\epsilon$ is the same as computed with the similar set of Yukawa couplings which are equal in magnitude but opposite in sign. Therefore in fig.\ref{fig4}, fig.\ref{fig5}, fig.\ref{fig6}, fig.\ref{fig7}, both the dark green and red regions are separated and show a symmetric pattern with respect to the origin (0,0). Here we want to clarify that from fig.\ref{fig4}, fig.\ref{fig5}, fig.\ref{fig6}, fig.\ref{fig7} it seems that the (0,0) points (where the real and imaginary parts of the Yukawa couplings are zero) are allowed by the constraints, which is misleading. For clarification, we would refer to fig.\ref{fig4}(b), which shows the zoomed version of fig.\ref{fig4}(a), where it is clearly seen that the (0,0) point is disallowed by the constraints. This conclusion is true for the other Yukawa couplings too.

 In fig.\ref{fig4}, fig.\ref{fig5}, fig.\ref{fig6}, fig.\ref{fig7}, the range of the real and imaginary part of Yukawa couplings filtered out by the neutrino oscillation data (represented by light green region) and constraint on lepton asymmetry (represented by dark green region), go on increasing with an increase in $\tan \beta$. Thus one can observe that after surviving through first two constraints, larger values of couplings become accessible for higher values of $\tan \beta$. For example, in the light green region in fig.\ref{fig4}, $y_{1R}$ ranges from $\sim -0.0035$ to $\sim 0.0035$ for $\tan \beta = 3$. Whereas, for $\tan \beta = 4$ and 5, the corresponding range extends to : $-0.0045 < y_{1R} < 0.0045$ and $-0.0058 < y_{1R} < 0.0058$ respectively. Likewise the imaginary part of $y_1$ behaves in a similar manner with increasing $\tan \beta$. The red regions do not follow the same pattern, {\em i.e.}
the red regions cover smaller Yukawa couplings as $\tan \beta$ grows in fig.\ref{fig4}, fig.\ref{fig5}, fig.\ref{fig6}, fig.\ref{fig7}. This observation holds for real and imaginary parts of the other couplings $y_{4}, y_{5}, y_{8}$ too (fig.\ref{fig5}, fig.\ref{fig6}, fig.\ref{fig7}). The shape of the distribution in fig.\ref{fig4}, fig.\ref{fig5} (elliptical) are different from that in fig.\ref{fig6}, fig.\ref{fig7} (quadrilateral). Sharp upper and lower edges in the $y_{5I} ~{\rm vs. }~ y_{5R}$ and $y_{8I} ~{\rm vs. }~ y_{8R}$ plane (fig.\ref{fig6}, fig.\ref{fig7}), parallel to $y_{5R}$ and $y_{8R}$ axes respectively, correspond to the perturbativity limits imposed on the Yukawa couplings, {\em i.e.} $|y_j| \leq \sqrt{4 \pi}$. Due to the perturbative requirement, upper and lower portions of the plots for $y_5$ and $y_8$ have been chopped off. Since the real and imaginary parts of $y_1$ and $y_4$ lie well within the perturbative limit already, the shape of the parameter space has no sharp edge in fig.\ref{fig4}, fig.\ref{fig5}. Here we have presented the plots for the Yukawa couplings $y_1, y_4, y_5, y_8$ only, because the shape of the plots for the rest of the Yukawa couplings $y_2, y_3, y_6, y_7, y_9$ resembles with that of the presented ones.

\section{Summary and conclusion}
\label{sec:6}
In this analysis, we have explored the possibility of neutrino mass generation via Type-I see-saw mechanism and baryogenesis via thermal leptogenesis in the  context of $Z_3$-symmetric 3HDM accompanied by three RH singlet neutrinos. According to the criteria of thermal leptogenesis, we consider hierarchical masses between three heavy neutrinos : $M_1 << M_2,~M_3$; $M_1$ having the lower limit of $10^9$ GeV. The thermal production and out-of-equilibrium decay of the lightest heavy neutrino $N_1$ gives rise to lepton asymmetry, which in turn is partially converted to baryon asymmetry via EW sphelaron processes. 

An important model parameter $\tan \beta$, relevant for lepton asymmetry calculation, has been filtered out by different constraints in the scalar sector. Among all of the constraints, the requirement of perturbativity of all quartic couplings, rules out the region of the parameter space with $\tan \beta > 5.5$. Thus we proceed with three discrete values of $\tan \beta$, {\em i.e.} $\tan \beta = 3, ~4, ~5$, to make a comparative study of the parameter space in the neutrino sector.


Among the three doublets, $\phi_3$ only being responsible for the neutrino mass generation due to $Z_3$-quantum number assignment, the Yukawa Lagrangian contains nine complex Yukawa couplings, {\em i.e.} 18 free parameters (real and imaginary parts of nine complex Yukawa couplings) to fit neutrino oscillation data, which is further simplified by CI parametrization. Three RH singlet neutrinos couple to the SM neutrinos via doublet $\phi_3$ only, and generate mass of light neutrinos via Type-I see-saw mechanism. From the decay of $N_1$, both lepton asymmetry $\epsilon$ and baryon asymmetry $Y_B$ are calculated at the points satisfying neutrino oscillation data with varying $M_1, M_2, M_3$. It is found that the available parameter space in $Y_B  -\epsilon$ plane shrinks with decreasing $\tan \beta$. To be consistent with the current bound on baryon asymmetry, one has to take $\epsilon \sim 10^{-6}$. Thus for the rest of the study, we have imposed a conservative limit of $1 \times 10^{-6} < |\epsilon| < 1.5 \times 10^{-6} $ on $\epsilon$ for all $\tan \beta$. This constraint immediately uplifts the lower bound on $M_1$ from $10^9$ GeV to $\sim 4 \times 10^{10} $ GeV, $\sim 2.5 \times 10^{10} $ GeV, $\sim 1.5 \times 10^{10} $ GeV for $\tan \beta = 3, ~4,~ 5$ respectively in our model. The available parameter space for a fixed $\tan \beta$ in the real vs. imaginary part of complex Yukawa coupling plane is diminished after applying three constraints sequentially : neutrino oscillation data, lepton asymmetry, baryon asymmetry. The last constraint turns out to be the most stringent among all.
\section{Acknowledgements}
Authors thank Dr. Joydeep Chakrabortty and Dr. Nabarun Chakrabarty for fruitful discussions. IC acknowledges support from DST, India, under grant number IFA18-PH214 (INSPIRE Faculty Award). HR is supported by the Science and Engineering Research Board, Government of India, under the agreement SERB/PHY/2016348 (Early Career Research Award).
\appendix

\section{\\ Decay width of $h \rightarrow \gamma \gamma$ in 3HDM}
\label{app : A}
Amplitude and decay width of the process $h \rightarrow \gamma \gamma$ can be written as \cite{Djouadi:2005gj}:
\besub
\bea
\mathcal{M}^{\text{3HDM}}_{h \to \gamma \gamma} &=& 
\sum_f N_f Q_f^2 f_{hff} A_{1/2}\Big(\frac{M^2_h}{4 M^2_f}\Big)
 + f_{h VV} A_1\Big(\frac{M^2_h}{4 M^2_W}\Big)\nonumber \\
&& 
 + \sum_{i=1}^2 \frac{\l_{h H_i^+ H_i^-} v}{2 M^2_{H_i^+}} A_0\Big(\frac{M^2_h}{4 M^2_{H_i^+}}\Big) \\ 
\Gamma^{\text{3HDM}}_{h \to \gamma \gamma} &=& \frac{G_F \a^2 M_h^3}{128 \sqrt{2} \pi^3} |\mathcal{M}^{\text{3HDM}}_{h \to \gamma \gamma}|^2,
\eea
\eesub
where $N_f, Q_f, G_F$ and $\a$ denote respectively color factor, charge of fermion, the Fermi constant and the QED fine-structure constant. For quarks $N_f = 3$. $\lambda_{h H_i^+ H_i^-}$ is $h H_i^+ H_i^-$ ($i = 1,2$) coupling. $f_{hff}, f_{hVV}$ are scale factors of $hff, hVV$ couplings with respect to SM. When the alignment limit is strictly enforced,
\bea
f_{hff} = f_{hVV} = 1
\eea

The loop functions are listed below.
\besub
\bea
A_{1/2}(x) &=& \frac{2}{x^2}\big((x + (x -1)f(x)\big), \\
A_1(x) &=& -\frac{1}{x^2}\big((2 x^2 + 3 x + 3(2 x -1)f(x)\big), \\
A_0(x) &=& -\frac{1}{x^2}\big(x - f(x)\big),  \\
\text{with} ~~f(x) &=& \text{arcsin}^2(\sqrt{x}); ~~~x \leq 1 
\nonumber \\
&&
= -\frac{1}{4}\Bigg[\text{log}\frac{1+\sqrt{1 - x^{-1}}}{1-\sqrt{1 - x^{-1}}} -i\pi\Bigg]^2; ~~~x > 1.
\eea
\eesub

where $A_{1/2}(x), A_1(x)$ and 
$A_0(x)$ are the respective amplitudes for the spin-$\frac{1}{2}$, spin-1 and spin-0 particles in the loop.

$\lambda_{h H_1^+ H_1^-}, ~\lambda_{h H_2^+ H_2^-}$ can be expressed in terms of quartic couplings and mixing angles as,

\bea
\lambda_{h H_1^+ H_1^-} & = & -\cos \beta_1 (\cos \alpha_3 (v \sin \beta_1 \cos \beta_2 \sin \gamma_2 \cos \gamma_2 (\lambda_9 \sin \beta_2+\lambda_{11}) \nonumber \\
&&   +\lambda_3 v \sin ^2 \beta_2 \sin ^2 \gamma_2 + \lambda_6 v \sin ^2 \beta_2 \cos ^2 \gamma_2) \nonumber \\ 
&& +\cos \beta_2
   (-\lambda_{12} v \sin \gamma_2 \cos \gamma_2 (\sin \alpha_3 \sin \beta_1 + \sin \beta_2) \nonumber \\
&&    +\lambda_2 v \sin \alpha_3 \sin
   \beta_1 \sin \beta_2 \cos ^2 \gamma_2 + \lambda_6 v \sin \alpha_3 \sin \beta_1 \sin \beta_2 \sin ^2 \gamma_2) \nonumber \\
   && + \lambda_9 v \sin \alpha_3 \sin ^2 \beta_2 \sin \gamma_2 \cos \gamma_2  
   - \lambda_{11} v \sin \beta_1 \cos ^2 \beta_2 \sin \gamma_2 \cos
   \gamma_2) \nonumber \\
   && +\cos ^2 \beta_1 (v \cos ^2 \beta_2 (\lambda_4 \cos ^2 \gamma_2 + \lambda_5 ~\sin ^2 \gamma_2)  \nonumber \\
&&    -\cos \beta_2 \sin \gamma_2 \cos \gamma_2 (\lambda_{11} v \sin \alpha_3 \sin \beta_2+\lambda_{12} v \cos \alpha_3 \sin \beta_2))  \nonumber \\
&&    +\sin
   \beta_1 (\sin \alpha_3 (\lambda_9 v \sin \beta_1 \cos \beta_2 \sin \gamma_2 \cos \gamma_2 +\lambda_3 v \sin
   \beta_2 \sin ^2 \gamma_2  \nonumber \\
&&    + \lambda_6 v \sin \beta_2 \cos ^2 \gamma_2)-\cos \alpha_3 (\cos \beta_2 (\lambda_2 v \sin
   \beta_1 \cos ^2 \gamma_2  \nonumber \\
 &&   + \lambda_6~ v \sin \beta_1 \sin ^2 \gamma_2)+\lambda_9 v \sin \beta_2 \sin \gamma_2 \cos
   \gamma_2))
\eea
\bea
\lambda_{h H_2^+ H_2^-} & = &-\cos \beta_1 (\sin \gamma_2 \cos \gamma_2(\cos \beta_2 (\lambda_{12} v \sin \alpha_3 \sin \beta_1+\lambda_{12} v \sin
   \beta_2) \nonumber \\
   && -\lambda_9 v \sin \alpha_3 \sin ^2\beta_2+\lambda_{11} v \sin \beta_1 \cos ^2\beta_2) \nonumber \\
   &&+\cos \alpha_3 (-\cos
   \beta_2 \sin \gamma_2 \cos \gamma_2 (\lambda_9 v \sin \beta_1 \sin \beta_2  \nonumber \\
   && +\lambda_{11} v \sin \beta_1)+\lambda_3 v \sin
   ^2\beta_2 \cos ^2\gamma_2+\lambda_6 v \sin ^2\beta_2 \sin ^2\gamma_2) \nonumber \\
   && +\lambda_2 v \sin \alpha_3 \sin \beta_1 \sin
   \beta_2 \cos \beta_2 \sin ^2\gamma_2 \nonumber \\
 &&   +\lambda_6 v \sin \alpha_3 \sin \beta_1 \sin \beta_2 \cos \beta_2 \cos ^2 \gamma_2)\nonumber \\
   && +\frac{1}{4} \cos ^2\beta_1 (2 \cos \beta_2 \sin 2\gamma_2 (\lambda_{11} v \sin \alpha_3 \sin \beta_2 \nonumber \\
   && +\lambda_{12} v \cos
   \alpha_3 \sin \beta_2)+2 v \cos ^2\beta_2 (\cos 2\gamma_2 (\lambda_5-\lambda_4)+\lambda_4+\lambda_5))\nonumber \\ 
   && +\sin \beta_1
   (\sin ^2\gamma_2 (\lambda_6 v \sin \alpha_3 \sin \beta_2-\lambda_2 v \cos \alpha_3 \sin \beta_1 \cos \beta_2) \nonumber \\
   && +\cos
   ^2\gamma_2 (\lambda_3 v \sin \alpha_3 \sin \beta_2
    -\lambda_6 v \cos \alpha_3 \sin \beta_1 \cos \beta_2) \nonumber \\
    && +\sin \gamma_2
   \cos \gamma_2 (\lambda_9 v \cos \alpha_3 \sin \beta_2 
   -\lambda_9 v \sin \alpha_3 \sin \beta_1 \cos \beta_2))
\eea

\section{Formulas for reduced cross sections}
\label{app : B}
Expression for $\gamma_{D_1}$ can be written as \cite{Plumacher:1996kc} :
\bea
\gamma_{D_1} = \gamma_{eq} = N_{N_1}^{eq}~ \frac{K_{1}(z)}{K_{2}(z)}~ \Gamma_{N_1} \,, {\rm with}~
 z = \frac{M_1}{T} \nonumber 
\eea
$N_{N_1}^{eq}$ being the equilibrium number density of the lightest RH neutrino $N_1$. Here $K_1$ and $K_2$ are the first and second modified Bessel functions of second kind respectively and $\Gamma_{N_1}$ is the total decay width of $N_1$.

For decay of $N_1$, $\gamma_{eq}$ can be written as \cite{Plumacher:1996kc}, 
\bea
\gamma_{eq} = \frac{T}{64 \pi^{4}} \int_{M^{2}_{1}}^{\infty} \hspace{1mm} ds \hat{\sigma}(s) \hspace{1mm} \sqrt{s} \hspace{1mm}  K_{1}(\frac{\sqrt{s}}{T}) \,.
\eea
where $s$ \footnote{Not to be confused with "$s$-channel" mentioned earlier.} is the square of center of mass energy and $\hat{\sigma}(s)$ is reduced cross section, which can be expressed in terms of actual cross section for two body scattering $ a + b \rightarrow i + j + ...$ as \cite{Plumacher:1996kc} :
\bea
\hat{\sigma}(s) = \frac{8}{s}\left[(p_a.p_b)^2 - M_a^2 M_b^2\right] \sigma(s) \,,
\eea
with $p_k$ and $M_k$ being three momentum and mass of particle $k$.

Decay width of $N_1$ at tree level, 
\bea
\Gamma_{N_1} &:= & \Gamma (N_1 \rightarrow \phi_3^{\dagger} + l) + \Gamma (N_1 \rightarrow \phi_3 + \bar{l}) \nonumber \\
&& = \frac{\alpha}{\text{sin}^{2}\theta_W} \frac{M_{1}}{4} \frac{(M^{\dagger}_{D} M_{D})_{11}}{M^{2}_{W}}
\label{decal-width}
\eea
with $\alpha, \theta_W$ being the Fine structure constant and the Weinberg angle.

The reduced cross-section of $N_{1}$ decay is given by \cite{Plumacher:1996kc},
\bea
\hat{\sigma}_{N_{1}}(s)&=  &\frac{\alpha^{2}}{\text{sin}^{4}\theta_W} \frac{2 \pi}{M^{4}_{W}}\frac{1}{x} a_{1}(M^{\dagger}_{D} M_{D})_{11}^2 \Big[\frac{x}{a_{1}} + \frac{2 x}{D_{1}(x)} + \frac{x^{2}}{2 D^{2}_{1}(x)} \nonumber \\ && - \Big(1 + 2\frac{x + a_{1}}{D_{1}(x)}\Big) ~\text{ln}\Big(\frac{x + a_{1}}{a_{1}}\Big) \Big],
\label{n1-decay}
\eea
where, $x = \frac{s}{M_1^2}, ~ a_1 =1, ~
\frac{1}{D_{1}(x)} := \frac{x - a_{1}}{(x-a_{1})^{2} + a_{1} c_{1}}$, with $c_{1} := \Big(\frac{\Gamma_{N_{1}}}{M_{1}}\Big)^{2}$
\newline
 The reduced cross-section for $L$-violating t-channel process (via $N_{1}$) is \cite{Plumacher:1996kc},
 \bea 
 \hat{\sigma}_{N_{1},t} (s) =  \frac{2 \pi \alpha^2 a_1}{M^{4}_{W} \text{sin}^{4}\theta} (M^{\dagger}_{D} M_{D})^{2}_{11}  \times \Big[\frac{1}{2 a_1}~\frac{x}{x+a_{1}} + \frac{1}{x + 2 a_{1}} \text{ln}\Big(\frac{x + a_{1}}{a_{1}}\Big)\Big].
 \label{n1-tchannel}
 \eea
 The reduced cross-section for s-channel process $N_1 + l \rightarrow \bar{t} + q$ (mediated by $\phi_3$) is \footnote{Here "$s$" in the subscript of $\hat{\sigma}^{1}$ signifies "$s$"-channel process and the argument of $\hat{\sigma}^{1}$ represents center of mass energy.} \cite{Plumacher:1996kc},
 \bea
 \hat{\sigma}^{1}_{\phi,s}(s) = \frac{3 \pi \alpha^{2} M^{2}_{t}}{M^{4}_{W} \text{sin}^{4}\theta_W} (M^{\dagger}_{D} M_{D})_{11} \Big(\frac{x - a_{1}}{x}\Big)^{2}.
 \label{higgs-schannel}
 \eea
 The reduced cross-section for $t$-channel process $N_1 + t \rightarrow \bar{l} + q$ (mediated by $\phi_3$) is \cite{Plumacher:1996kc},
 \bea
 \hat{\sigma}^{1}_{\phi,t}(s) = \frac{3 \pi \alpha^{2} M^{2}_{t}}{M^{4}_{W} \text{sin}^{4}\theta_W} (M^{\dagger}_{D} M_{D})_{11} \times \Big[\frac{x - a_{1}}{x} + \frac{a_{1}}{x}~ \text{ln} \Big(\frac{x - a_{1} + y~'}{y~'}\Big)\Big],
 \label{higgs-tchannel}
 \eea
 where $y~' = \frac{M^{2}_h} {M^{2}_{1}}$.
\section{Expressions of $(Y^\dag Y)_{11},~ {\rm Im} \{(Y^\dag Y)_{12}^2\},~ {\rm Im} \{(Y^\dag Y)_{13}^2\}$}
\label{app:C}
Expressions for $(Y^\dag Y)_{11},~ {\rm Im} \{(Y^\dag Y)_{12}^2\},~ {\rm Im} \{(Y^\dag Y)_{13}^2\}$  in eq.(\ref{lepton_asy}) can be written in terms of real and imaginary parts of Yukawa couplings as :
\bea
(Y^\dag Y)_{11} &=& y_{1R}^2 + y_{1I}^2 + y_{4R}^2 + y_{4I}^2 + y_{7R}^2 + y_{7I}^2 \,. \nonumber \\
{\rm Im} \{(Y^\dag Y)_{12}^2\} &=& y_{1R}^2 y_{2R} y_{2I}-2 y_{1R} y_{1I} y_{2R}^2   
 +2 y_{1R}
   y_{1I} y_{2I}^2+2 y_{1R} y_{2R} y_{4R} y_{5I}-2 y_{1R}
   y_{2R} y_{4I} y_{5R} \nonumber \\
   &&+2 y_{1R} y_{2R} y_{7R} y_{8I}-2
   y_{1R} y_{2R} y_{7I} y_{8R}+2 y_{1R} y_{2I} y_{4R}
   y_{5R}+2 y_{1R} y_{2I} y_{4I} y_{5I} \nonumber \\
   && +2 y_{1R} y_{2I}
   y_{7R} y_{8R}+2 y_{1R} y_{2I} y_{7I} y_{8I}-2 y_{1I}^2
   y_{2R} y_{2I}-2 y_{1I} y_{2R} y_{4R} y_{5R}  \nonumber \\
   && -2 y_{1I}
   y_{2R} y_{4I} y_{5I}-2 y_{1I} y_{2R} y_{7R} y_{8R}
    -2
   y_{1I} y_{2R} y_{7I} y_{8I} 
   +2 y_{1I} y_{2I} y_{4R}
   y_{5I} \nonumber \\
   && -2 y_{1I} y_{2I} y_{4I} y_{5R}+2 y_{1I} y_{2I}
   y_{7R} y_{8I}-2 y_{1I} y_{2I} y_{7I} y_{8R}+2 y_{4R}^2
   y_{5R} y_{5I} \nonumber \\
   &&-2 y_{4R} y_{4I} y_{5R}^2+2 y_{4R} y_{4I}
   y_{5I}^2+2 y_{4R} y_{5R} y_{7R} y_{8I} 
   -2 y_{4R} y_{5R}
   y_{7I} y_{8R}+2 y_{4R} y_{5I} y_{7R} y_{8R} \nonumber \\
   && +2 y_{4R}
   y_{5I} y_{7I} y_{8I}-2 y_{4I}^2 y_{5R} y_{5I}-2 y_{4I}
   y_{5R} y_{7R} y_{8R}-2 y_{4I} y_{5R} y_{7I} y_{8I}+2
   y_{4I} y_{5I} y_{7R} y_{8I} \nonumber \\
   && -2 y_{4I} y_{5I} y_{7I}
   y_{8R}+2 y_{7R}^2 y_{8R} y_{8I}-2 y_{7R} y_{7I} y_{8R}^2+2
   y_{7R} y_{7I} y_{8I}^2-2 y_{7I}^2 y_{8R} y_{8I} \,. \nonumber \\
   {\rm Im} \{(Y^\dag Y)_{13}^2\} &=& 2 y_{1R}^2 y_{3R} y_{3I}-2 y_{1R} y_{1I} y_{3R}^2+2 y_{1R}
   y_{1I} y_{3I}^2+2 y_{1R} y_{3R} y_{4R} y_{6I} 
   -2 y_{1R}
   y_{3R} y_{4I} y_{6R} \nonumber \\
   && +2 y_{1R} y_{3R} y_{7R} y_{9I}-2
   y_{1R} y_{3R} y_{7I} y_{9R}+2 y_{1R} y_{3I} y_{4R}
   y_{6R}+2 y_{1R} y_{3I} y_{4I} y_{6I} \nonumber \\
   && +2 y_{1R} y_{3I}
   y_{7R} y_{9R}+2 y_{1R} y_{3I} y_{7I} y_{9I}-2 y_{1I}^2
   y_{3R} y_{3I}-2 y_{1I} y_{3R} y_{4R} y_{6R} \nonumber \\
   && -2 y_{1I}
   y_{3R} y_{4I} y_{6I}-2 y_{1I} y_{3R} y_{7R} y_{9R}-2
   y_{1I} y_{3R} y_{7I} y_{9I}+2 y_{1I} y_{3I} y_{4R}
   y_{6I} \nonumber \\
   && -2 y_{1I} y_{3I} y_{4I} y_{6R}+2 y_{1I} y_{3I}
   y_{7R} y_{9I} 
    -2 y_{1I} y_{3I} y_{7I} y_{9R}+2 y_{4R}^2
   y_{6R} y_{6I}-2 y_{4R} y_{4I} y_{6R}^2 \nonumber \\
   && +2 y_{4R} y_{4I}
   y_{6I}^2+2 y_{4R} y_{6R} y_{7R} y_{9I} 
      -2 y_{4R} y_{6R}
   y_{7I} y_{9R} 
   +2 y_{4R} y_{6I} y_{7R} y_{9R} \nonumber \\
   && +2 y_{4R}
   y_{6I} y_{7I} y_{9I}-2 y_{4I}^2 y_{6R} y_{6I} 
    -2 y_{4I}
   y_{6R} y_{7R} y_{9R} 
    -2 y_{4I} y_{6R} y_{7I} y_{9I} 
    +2
   y_{4I} y_{6I} y_{7R} y_{9I} \nonumber \\
   && -2 y_{4I} y_{6I} y_{7I}
   y_{9R}+2 y_{7R}^2 y_{9R} y_{9I}-2 y_{7R} y_{7I} y_{9R}^2+2
   y_{7R} y_{7I} y_{9I}^2-2 y_{7I}^2 y_{9R} y_{9I}.
   \label{epsi_part}
\eea   
\section{Computation of bosonic and fermionic degrees of freedom (D.O.F)}
\label{app:D}
For fermionic sector D.O.F can be computed as \cite{Plehn:2017fdg},

\bea
g_{\text{fermion}} &=& g_{\text{quark}} + g_{\text{lepton}} + g_{\text{neutrino}} + g_{\text{RH-neutrino}} \nonumber \\
&&=  (6 \times 3\times 2 \times 2) + (3 \times 2 \times 2) + (3 \times 2 + 3 \times 2) = 96 \,.
 \label{effective-dof_fermion}
 \eea
 Similarly for the bosonic counter part, D.O.F can be calculated as \cite{Plehn:2017fdg}, 
 \bea
g_{\text{boson}} &=& g_{\text{gluon}} + g_{\text{weak}} + g_{\text{photon}} + g_{\text{Higgs}} + g_{\text{BSM-Higgs}} \nonumber \\
&&= (8 \times 2) + (3 \times 3) + 2 + 1 + 8 = 36  \,.
\label{effective-dof_boson}    
\eea
For 3HDM the total effective D.O.F can be obtained by summing up the D.O.Fs in fermion and bosonic sectors,

\bea
g_{eff} (T > 174~ {\rm GeV}) = 36 + \frac{7}{8} \times 96 = 120.
\label{eff-dof-3HDM}
\eea 
\bibliography{refer1}
\end{document}